%% file: main.tex
\documentclass[journal,draftclsnofoot,onecolumn,12pt,twoside]{IEEEtran}
\usepackage{todonotes}

\usepackage{cite}
\usepackage{xcolor}
\usepackage{epsfig}
\usepackage{textcomp}
\usepackage{indentfirst}
\usepackage{xcolor}
\usepackage{times}
\usepackage{float}
\usepackage{algorithm}
\usepackage{algpseudocode}
\usepackage{amsmath,amssymb,amsfonts}
\usepackage{afterpage}
\usepackage{sansmath}
\usepackage{amsmath}
\usepackage{amstext,cite}
\usepackage{amssymb,bm}
\usepackage{latexsym}
\usepackage{color}
\usepackage{graphicx}
\usepackage{amsmath}
\usepackage{amsthm}
\usepackage{graphicx}
\usepackage[center]{caption}
\usepackage{pstricks}
\usepackage{pdfpages}
\usepackage[font=footnotesize,labelfont=bf]{caption}

\usepackage{subcaption}
\usepackage{booktabs}
\usepackage{multicol}
\usepackage{lipsum}% dummy text
\usepackage[T1]{fontenc}
\usepackage{enumitem}
 \usepackage{aecompl}
 \usepackage{mathrsfs}
\usepackage{setspace}
\usepackage{amssymb} 
\usepackage{caption}
\usepackage{arydshln}
\usepackage{soul}
\usepackage{cancel}
\usepackage{changes}
\definechangesauthor[color=red,name={Mingyue Ji}]{MJ}
\def\BibTeX{{\rm B\kern-.05em{\sc i\kern-.025em b}\kern-.08em
    T\kern-.1667em\lower.7ex\hbox{E}\kern-.125emX}}
\allowdisplaybreaks

\input{macros}

\theoremstyle{definition}

\newtheorem{proposition}{Proposition}
\theoremstyle{definition}
\newtheorem{definition}{Definition}
\newtheorem{construction}{Construction}

\newtheorem{remark}{Remark}
\newtheorem{ex}{Example}
\begin{document}

\title{Reconfigurable Intelligent Surface-Assisted Multiple-Antenna  Coded Caching\\
}

\author{Xiaofan~Niu, Minquan~Cheng,  Kai~Wan,~\IEEEmembership{Member,~IEEE,} Robert~Caiming~Qiu,~\IEEEmembership{Fellow,~IEEE}
and~Giuseppe~Caire,~\IEEEmembership{Fellow,~IEEE}

\thanks{
A short version of this paper   was presented at the 2024 IEEE Information Theory Workshop (ITW)~\cite{niu2024reflecting}.
}
\thanks{X.~Niu,  K.~Wan, and R.~C.~Qiu are with the School of Electronic Information and Communications,
Huazhong University of Science and Technology, 430074  Wuhan, China,  (e-mail: \{nxfan, kai\_wan,caiming\}@hust.edu.cn). The work of X.~Niu,  K.~Wan, and R.~C.~Qiu  was partially funded by the   National Natural
Science Foundation of China (NSFC-12141107),  the Key Research and Development Program of Wuhan under Grant 2024050702030100, and Wuhan ``Chen
Guang'' Pragram under Grant 2024040801020211.}
\thanks{M. Cheng is with Guangxi Key Lab of Multi-source Information Mining $\&$ Security, Guangxi Normal University,
Guilin 541004, China  (e-mail:  chengqinshi@hotmail.com). The work of M.~Cheng was partially funded by 2022GXNSF (No.A035087, BA035616), 2024GXNSFGA010001,  and the BAGUI Young Scholar Program of Guangxi.}
\thanks{G. Caire is with the Electrical Engineering and Computer Science Department, Technische Universit\"{a}t Berlin,
10587 Berlin, Germany (e-mail: caire@tu-berlin.de). 
The work of G. Caire was partially funded by the %BMBF Germany in the program of ``Souverän. Digital. Vernetzt.'' Joint Project 6G-RIC (Project IDs 16KISK030).
Gottfried Wihelm Leibniz-Preis 2021 of the German Science Foundation (DFG). 
}
}

\maketitle

\begin{abstract}
Reconfigurable Intelligent Surface (RIS) has emerged as a promising technology to enhance the wireless propagation environment for next-generation wireless communication systems. 
This paper introduces a new RIS-assisted multiple-antenna coded caching problem. Unlike the existing multi-antenna coded caching models, our considered model incorporates a passive RIS with a limited number of elements aimed at enhancing the multicast gain (i.e., Degrees of Freedom (DoF)). The system consists of a server equipped with multiple antennas and several single-antenna users. The RIS, which functions as a passive and configurable relay, improves communication by selectively 'erasing' certain transmission paths between transmit and receive antennas, thereby reducing interference. 
We first propose a new RIS-assisted interference nulling algorithm to determine the phase-shift coefficients of the RIS. This algorithm achieves  faster convergence compared to the existing approach. By strategically nulling certain interference paths in each time slot, the transmission process is divided into multiple interference-free groups. Each group consists of a set of transmit antennas that serve a corresponding set of users without any interference from other groups. The optimal grouping strategy to maximize the DoF is formulated as a combinatorial optimization problem. To efficiently solve this, we design a low-complexity algorithm that identifies the optimal solution and develops a corresponding coded caching scheme to achieve the maximum DoF. 
Building on the optimal grouping strategy, we introduce a new framework, referred to as   \textit{RIS-assisted Multiple-Antenna Placement Delivery Array (RMAPDA)}, to construct the cache placement and delivery phases. Then we propose a general RMAPDA design  to achieve the maximum DoF under the optimal grouping strategy. 
In summary, this paper pioneers the integration of RIS into coded caching systems to boost multicast gain, offering a new direction for the future of wireless communications.
\end{abstract}


\section{Introduction}
\subsection{Coded Caching}
Coded caching, first introduced by Maddah-Ali and Niesen (MN) in~\cite{maddah2014fundamental}, has emerged as a powerful technique to alleviate network congestion. Beyond providing local caching gain—where users prefetch content to reduce server load—coded caching exploits multicast opportunities by leveraging the overlap in cached content among users. This enables a single transmission to simultaneously serve multiple users, thereby significantly reducing network traffic.

The original coded caching problem was formulated in a single-input single-output (SISO) shared-link model, where a central server with access to a library of $N$ files serves $K$ users via a noiseless shared link, and each user is equipped with a cache that can store $M$ files. A typical coded caching scheme involves two phases: a \textit{placement phase}, during which users independently cache content without knowledge of future demands, and a \textit{delivery phase}, where the server transmits coded multicast messages based on the users’ requests and their cached contents, aiming to minimize the worst-case load.

Under the MN scheme, each file is split into subfiles such that each subfile is cached by $t = KM/N \in \{0, 1, \ldots, K\}$ users. In the delivery phase, each multicast transmission serves $t+1$ users, yielding a coded multicast gain of $t+1$. The corresponding load is $(K - t)/(t + 1)$, where the numerator reflects local caching gain and the denominator reflects multicast gain. It has been shown in~\cite{indexcodingcaching2020,exactrateuncoded} that, under the constraint of uncoded cache placement, the MN scheme is optimal in terms of coded caching gain when $N \geq K$.

Despite its theoretical elegance, the MN scheme suffers from high subpacketization, which severely limits its practical deployment. To address this, a combinatorial framework known as the \textit{Placement Delivery Array} (PDA) was introduced in~\cite{yan2017placement}. PDAs provide a structured way to design coded caching schemes with uncoded cache placement and clique-covering delivery. Various combinatorial techniques, such as injective arc coloring on digraphs~\cite{wu2023design}, hypergraphs, bipartite graphs, projection spaces, and block designs, have been employed to reduce subpacketization while preserving multicast gains~\cite{tang2018coded,shangguan2018centralized,shanmugam2017coded,cheng2021linear,krishnan2018coded,yan2017placement,wang2025pda}.

The MN framework has also been extended to more complex wireless networks, including Device-to-Device (D2D) networks~\cite{li2022new}, hierarchical caching architectures~\cite{karamchandani2016hierarchical}, and linear multi-server systems~\cite{ji2015fundamental, ji2015combination, shariatpanahi2016multi}. It has further been applied to interference-limited settings with multiple single-antenna transmitters and receivers, aiming to maximize the sum Degrees of Freedom (sum-DoF)~\cite{naderializadeh2017fundamental,hachem2018degrees}. When all transmitters have access to the full file library, the system reduces to a cache-aided multiple-input single-output (MISO) broadcast channel (BC)~\cite{shariatpanahi2018physical,lampiris2018adding,salehi2021low,mohajer2020miso,salehi2020diagonal,salehi2019subpacketization,piovano2019generalized}. Under one-shot linear delivery combining cache placement and zero-forcing (ZF) precoding, the achievable sum-DoF is given by $\min\left\{L + \frac{KM}{N}, K\right\}$, where $L$ is the number of transmit antennas~\cite{naderializadeh2017fundamental,lampiris2021resolving}. $\min\left\{L + \frac{KM}{N}, K\right\}$ was then proved to be the optimal sum-DoF under the constraint of uncoded cache placement and one-shot linear delivery. 
However, a major bottleneck in MISO coded caching is again the subpacketization, which is much larger than that of the MN scheme. %In some cases, it scales as $\binom{K}{KM/N} \cdot \binom{K - KM/N}{L - 1}$~\cite{shariatpanahi2016multi,shariatpanahi2018physical}, posing serious limitations. 
 To mitigate this, several approximate constructions and grouping strategies have been proposed~\cite{lampiris2018adding,salehi2021low,mohajer2020miso,salehi2020diagonal,salehi2019subpacketization}. 
 \iffalse 
 When both $K/L$ and $KM/(LN)$ are integers, the subpacketization can be reduced to $\binom{K/L}{KM/(LN)}$~\cite{lampiris2018adding}, while in general scenarios, near-optimal designs with bounded DoF gaps (e.g., $5/3$ for $L>t$ and $4/3$ for $L<t$) have been developed, with the gap vanishing as $K$ increases. Moreover, when $L \geq t$, cyclic cache placement enables constructions with subpacketization scaling linearly in $K$~\cite{salehi2021low}.
\fi 
To systematically address the subpacketization challenge in multi-antenna settings, PDA  has been extended to the cache-aided MISO systems, where the resulting framework is referred to as \textit{Multiple-Antenna Placement Delivery Array} (MAPDA)~\cite{yang2023multiple,namboodiri2023extended}. MAPDA integrates spatial multiplexing and interference management into the original PDA combinatorial design, enabling the design of MISO coded caching schemes with one-shot ZF-based delivery schemes. A MAPDA was proposed in~\cite{yang2023multiple} which results in a MISO coded caching scheme with the optimal sum-DoF $\min\left\{L + \frac{KM}{N}, K\right\}$, but with the same order of subpacketization as the original MN scheme. 
Subsequent advances include the MAPDA constructions with cyclic cache placement~\cite{wan2022multiple}, and the extensions to more complicated network topologies such as  % and hierarchical extensions~\cite{pandey2024coded}, which aim to maintain scalability and efficiency. To further broaden applicability, MAPDA-based schemes have been generalized to 
MISO coded caching schemes with partial connection~\cite{cheng2024coded}, multi-access networks~\cite{peter2023multi,cheng2024design}. %Two-dimensional designs incorporating cyclic wrap-around and orthogonal access modes have also been proposed~\cite{zhang2023coded,namboodiri2024two}. For heterogeneous cache scenarios, asymmetric storage capacities have been considered~\cite{mahmoodi2023multi}.

In addition, recent works have focused on improving decoding efficiency and deployment practicality. SIC-free delivery strategies~\cite{sojdeh2025sic}, multicast group assignment and beamforming~\cite{gamaethige2022reduced}, and weighted beamforming designs~\cite{mahmoodi2024low} have been introduced to reduce latency and simplify receiver processing. Furthermore, location-aware and mobility-robust schemes have been developed to accommodate dynamic user behaviors~\cite{abolpour2024cache}.

Collectively, these efforts provide a comprehensive foundation for the advancement of practical, scalable, and high-throughput multi-antenna coded caching systems under realistic wireless network constraints.

\subsection{Reconfigurable Intelligent Surface (RIS)}
\begin{figure} 
\centering
\includegraphics[width=0.8\linewidth, height=0.45\linewidth]{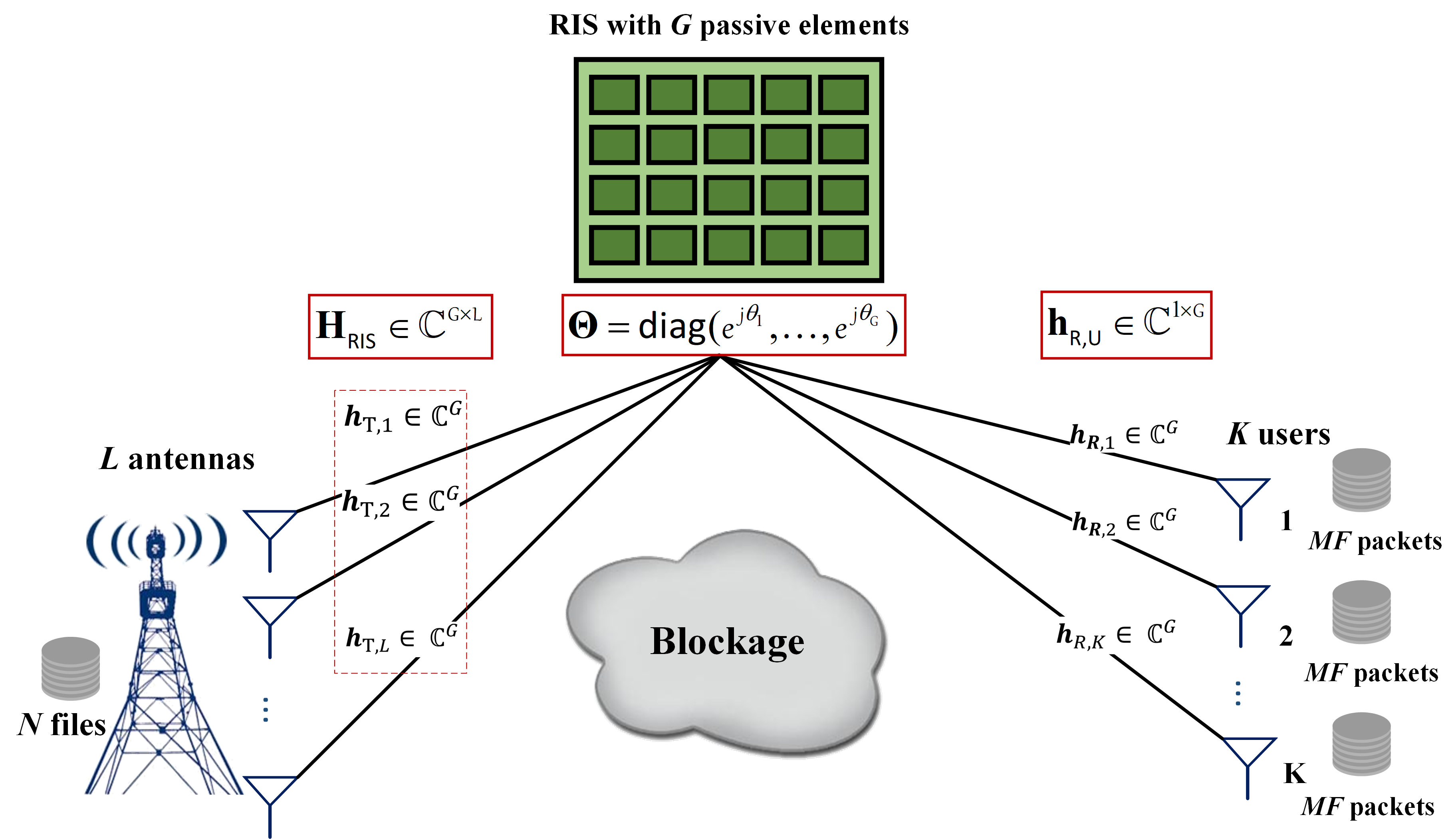} 
\caption{RIS-assisted multi-antenna coded caching system.}
\label{fig:system model}
\end{figure}

In this paper, we formulate a new coded caching system in wireless channels by  
introducing Reconfigurable Intelligent Surface (RIS)  into multi-antenna coded caching systems to further enhance the sum-DoF, where RIS is an advanced technology for next-generation wireless systems to reconfigure the wireless propagation environment~\cite{cui2014coding,8910627,di2020smart,9475160,liu2021reconfigurable}. 

Acting as a low-cost and passive beamformer, RIS, which was first introduced in 2010~\cite{zheludev2010road}, 
can intelligently reconfigure the wireless environment by dynamically controlling the phase shifts of reflected signals~\cite{liu2021reconfigurable}. This enables an improved signal-to-noise ratio (SNR), extended coverage, and enhanced link reliability without additional power consumption, aligning well with the dual goals of ultra-high data rates and energy efficiency in 6G networks.
RIS is an artificial planar structure composed of engineered electromagnetic materials and is electronically controlled by integrated circuits~\cite{sharma2021reconfigurable}. By manipulating the response of electromagnetic waves, RIS enables the dynamic reconfiguration of the wireless propagation environment, thereby enhancing channel gain, improving the quality of service (QoS), and extending coverage—all without incurring additional power consumption. Beyond these advantages, RIS also offers functionalities such as spatial blind spot mitigation, edge coverage enhancement, and inter-cell interference suppression. Furthermore, RIS can be integrated with various emerging technologies to enable a wide range of innovative applications. Owing to these distinctive features, RIS is widely regarded as a key enabling technology for future wireless networks, particularly 6G.

As illustrated in Fig.~\ref{fig:system model}, a widely adopted modeling approach for RIS was originally introduced by Wu and Zhang~\cite{wu2019intelligent}, which characterizes a RIS with $G$ reflecting elements via a diagonal phase-shift matrix $\boldsymbol{\Theta} = \mathrm{diag}(e^{j\theta_1}, \dots, e^{j\theta_G})$, where each diagonal entry represents the controllable phase of an individual reflecting unit. Under this model, the RIS-assisted channel is described as a cascaded composition of the BS-to-RIS channel $\mathbf{H}_{\text{RIS}} \in \mathbb{C}^{G \times L}$, the RIS phase-shift matrix $\boldsymbol{\Theta}$, and the RIS-to-user channel $\mathbf{h}_{\text{RU}}^H \in \mathbb{C}^{1 \times G}$. The resulting end-to-end effective channel can be expressed as
\begin{equation}
    \mathbf{h}_{\text{eff}}^H = \mathbf{h}_{\text{RU}}^H \boldsymbol{\Theta} \mathbf{H}_{\text{RIS}}.
\end{equation}
This modeling framework, which is also adopted in this work, enables the joint design of active beamforming at the transmitter and passive beamforming at the RIS, and serves as a fundamental basis for RIS-assisted wireless communication system design~\cite{wu2019intelligent,huang2019reconfigurable}, significantly improving energy efficiency in multi-user   cellular networks by jointly optimizing transmit power and reflection coefficients.  

RIS has also emerged as a powerful tool for interference management, which is critical in capacity-limited wireless environments. Zhang {\it et al.}~\cite{zhang2020capacity} analyzed the capacity limits of RIS-assisted point-to-point MIMO systems, and Nadeem and Alwazani~\cite{alwazani2020intelligent} studied the SINR maximization under power constraints—implicitly demonstrating RIS’s ability to mitigate interference by reconfiguring the wireless propagation environment.  Please refer to the surveys for more information on RIS~\cite{kaur2024survey, umer2025reconfigurable, basar2019wireless, bjornson2022reconfigurable, pan2022overview}.

Information theoretic works on  RIS mainly focus on using RIS to improve the system Degrees of Freedom (DoF).  
Bafghi {\it et al.}~\cite{bafghi2020degrees} investigated RIS-assisted $K$-user interference channels and proposed an interference alignment scheme that increases the achievable sum-DoF from $K/2$ to $K$ with appropriate RIS configurations. Extending this line of work, Jiang {\it et al.}~\cite{jiang2022interference} introduced a RIS-based interference nulling method combined with zero-forcing beamforming, achieving full DoF when the number of RIS elements exceeds a critical threshold on the order of $2K(K-1)$. An alternating projection algorithm was proposed in~\cite{jiang2022interference} to 
optimize the RIS phase-shift matrix for interference nulling.
%These interference management techniques have also been extended to RIS-assisted MIMO systems. 
Fu {\it et al.}~\cite{fu2021reconfigurable} studied interference alignment in RIS-assisted MIMO device-to-device (D2D) networks and proposed a low-rank joint transceiver and RIS design framework based on manifold optimization. By simultaneously optimizing the RIS phase-shift matrix and the transceiver beamforming matrices, this approach improves the system DoF. 

\subsection{Main Contribution}  
In order to further improve the sum-DoF in the multi-antenna coded caching systems, 
this paper is  the first to systematically integrate RIS into the multi-antenna coded caching framework.\footnote{\label{foot:active RIS}Recently, after the conference version of this paper, 
a work on active RIS-assisted coded caching was proposed in~\cite{changizi2024degrees}, 
 which focuses on DoF enhancement via uncoded caching and interference alignment in single-antenna interference channels assisted by active RISs. The main differences between our results and the results in~\cite{changizi2024degrees} include: (i) passive RIS v.s. active RIS; (ii) the coded caching schemes are totally different, where with the varying numbers of RIS elements, our schemes can achieve a significantly wider range of DoFs; (iii) we propose  a general combinatorial structure to construct  RIS-assisted coded caching schemes under uncoded cache placement and one-shot ZF-based delivery.} %{\red SUMMARIZE THE MODEL}   
We consider a RIS-assisted multi-antenna coded caching system with the following components: 

\begin{itemize}
    \item A centralized \textbf{server} equipped with $L$ transmit antennas and access to a library of $N$ files, each of length $F$ packets;
    \item $K$ single-antenna \textbf{users}, each with a local cache capable of storing up to $MF$ packets;
    % where $F$ denotes the subpacketization level
    \item A passive \textbf{RIS} with $G$ reflecting elements, which dynamically reconfigures the wireless propagation environment to facilitate interference mitigation.
\end{itemize}

The system operates in two phases: 

\textbf{Placement Phase:} Each file is partitioned into $F$ packets, which are stored across user caches without knowledge of future demands. The placement strategy can be either deterministic or random. 

\textbf{Delivery Phase:} After each user requests a file from the library, the server delivers the requested packets over a finite number of time slots using a \emph{one-shot linear delivery} scheme. Specifically, each packet is transmitted without symbol extension, and linear precoding is applied across the transmit antennas. The RIS is reconfigured in each time slot to assist in suppressing inter-user interference.

We assume that perfect Channel State Information (CSI) is available at both the server and the users. All transmissions are assumed to occur via RIS reflection, with the direct paths between the server and the users blocked. Nonetheless, the proposed RIS-assisted interference management framework is also applicable to systems with direct links, as demonstrated in prior work.

The main objective of this study is to develop cache placement and delivery strategies that \textbf{maximize the one-shot linear sum-DoF}, defined as the number of users that can be simultaneously served without interference in a single transmission round. As a high-SNR metric, the sum-DoF offers a tractable and insightful characterization of the performance of spatial and reflection-domain precoding in multi-user cache-aided MIMO networks.
 
 Besides the new problem formulation, the main contribution of this paper is summarized as follows:
\begin{itemize}  

    \item\textbf{New RIS-assisted interference nulling algorithm. }
   We propose a new RIS-assisted interference nulling algorithm that effectively `eliminates' some designated interference paths 
   from the transmitters to the users by incorporating local geometric refinement via tangent-space projection. 
Inspired by recent advances in Riemannian optimization~\cite{wei2016guarantees}, the proposed algorithm exploits the intrinsic structure of the feasible set by projecting the search direction onto the tangent space of the unit-modulus constraint manifold at each iteration. As a result, 
     the proposed algorithm efficiently obtains the RIS phase-shift coefficients and achieves significantly faster convergence compared to the alternating projection method in~\cite{jiang2022interference}.
     
    \item\textbf{Three-step coded caching construction.} 
    To achieve the optimal one-shot linear sum-DoF in a RIS-assisted multi-antenna coded caching system, we propose a structured three-step design framework that systematically integrates RIS-enabled interference management with coded caching techniques.

\begin{enumerate}
    \item \textbf{RIS-assisted grouping.} 
    We leverage the interference nulling capability of the RIS to partition each time slot in the delivery phase into multiple independent groups of transmit antennas and users. Within each group, users are served without interference from other groups. To maximize the achievable sum-DoF, the grouping problem is formulated as a combinatorial optimization task, and a low-complexity heuristic algorithm is developed to identify an efficient grouping strategy.
    
    \item \textbf{RMAPDA structure design.} 
    Given the optimal grouping strategy, we introduce a novel combinatorial structure termed \emph{RMAPDA} (RIS-assisted Multi-Antenna Placement Delivery Array), which generalizes the MAPDA framework~\cite{yang2023multiple,namboodiri2023extended} by incorporating RIS-enabled path elimination. This structure serves as a unified design framework for interference-aware cache placement and content delivery.

    \item \textbf{RMAPDA construction scheme.} By combining SMK–MAPDA and MN–PDA, and aligning time slots via perfect matching on a regular bipartite graph, we design an explicit $(L_0,K,F,Z,r,S)$ RMAPDA under $t<K$ for $t=KM/N$. Building on this, for any $L\ge L_0$ we obtain a RIS-assisted coded-caching scheme that achieves the optimal sum-DoF $\min \{K,L_0+tr\}$ at memory ratio $M/N=t/K$, with subpacketization $F$ and the number of RIS elements $G=2(r-1)((t+2)L_0-r)$. 
\end{enumerate}

\end{itemize}  

\paragraph*{Notation Convention}  
We represent scalars by lowercase letters, vectors by bold lowercase letters, and matrices by bold uppercase letters. For a matrix $\bm{A}$, $\bm{A}^T$ denotes its transpose, $\bm{A}^*$ denotes its conjugate, and $\bm A^{\mathrm H}$ denotes the conjugate transpose (Hermitian), i.e.,
$\bm A^{\mathrm H} := (\bm A^*)^T = (\bm A^T)^*$. The magnitude of a complex scalar $v$ is denoted by $|v|$. The operators $\text{Re}[\cdot]$ and $\text{Im}[\cdot]$ extract the real and imaginary parts of a complex number, respectively. The symbol ``$\cdot$'' indicates element-wise matrix multiplication, while $\odot$ represents the Hadamard product.  $\text{lcm}(a, b)$ denotes the least common multiple of $a$ and $b$. 
For a vector $\mathbf{v}$, $\text{diag}(\mathbf{v})$ denotes the diagonal matrix with the entries of $\mathbf{v}$ on its diagonal. We define $[a] = \{1, 2, \dots, a\}$. For a set $\Xc$ with $|\Xc| \geq y > 0$, $\binom{\Xc}{y}$ represents all subsets of $\Xc$ of size $y$. For integers $a$ and $b\ge 1$, $a \bmod b$ denotes the non-negative remainder of $a$ divided by $b$, taking values in $\{0,1,\dots,b-1\}$. Finally, $\mathbf{P}(i, j)$ represents the element in the $i$-th row and $j$-th column of the array $\mathbf{P}$.

\section{System Model and Related Results}
\subsection{System model}
As shown in Fig.~\ref{fig:system model}, we consider a $(L, G, K, M, N)$ RIS-assisted multi-antenna coded caching system.  
The server, equipped with $L$  antennas, has access to a library of $N$ files, denoted by $\mathcal{W} = \{\mathbf{W}_n \mid n \in [N]\}$, where each file $\mathbf{W}_n$ consists of $F$ packets, i.e., $\mathbf{W}_n \triangleq \{\mathbf{W}_{n,f} \mid f \in [F]\}$. Each user, equipped with a single antenna, has a cache capable of storing up to $MF$ packets, where $0 \leq M \leq N$.

A $(L,G,K,M,N)$ coded caching scheme in this system consists of two distinct phases:

\textbf{Placement Phase:} The $F$ packets of each file are strategically stored in users' caches without prior knowledge of their future requests.

\textbf{Delivery Phase:}  Each user $k \in [K]$ requests a file $\mathbf{W}_{d_k}$ from the library, where $d_k \in [N]$. The request vector is denoted as $\bm{d} = (d_1, d_2, \dots, d_K)$. Each packet is encoded using Gaussian encoding at a rate of $\log P$, where $P$ is the Signal-to-Noise Ratio (SNR). For sufficiently large $P$, each coded packet conveys one DoF. \footnotetext{
The \emph{Degrees of Freedom} (DoF) intuitively represents the number of interference-free data streams that can be simultaneously transmitted in the high-SNR regime. More precisely, consider a Gaussian point-to-point channel with an input power constraint \(P\), where signals are transmitted using Gaussian codebooks at a rate of \(\log P + o(\log P)\). The DoF is defined as the pre-log factor of the achievable rate, given by $\lim_{P \to \infty} \frac{R}{\log P}$, which evaluates to 1 in this setting. Therefore, in the high-SNR regime, each coded packet transmitted at rate \(\log P\) can be regarded as carrying one DoF.}

The encoded packets are represented as $\widetilde{\mathbf{W}}_{n,f}$, where $n\in [N]$ and $f\in [F]$.

The  one-shot linear delivery as in~\cite{naderializadeh2017fundamental} is considered. More precisely,  
the communication process contains $S$ time slots. During each time slot $s \in [S]$, we define the service set as a set of pairs (user index, packet index) 
\[ \mathcal{K}_s \;\triangleq\; \{(k_1^s,f_1^s),\,(k_2^s,f_2^s),\,\ldots,\,(k_{r_s}^s,f_{r_s}^s)\}, 
\] 
intended for a set of $r_s$ users. For each $i\in [L]$, the signal transmitted by the $i$-th transmitter (i.e., the $i$-th antenna) is given by:
\[
X_i(s) = \sum_{(k,f) \in \mathcal{K}_s} m_{i,k}(s) \widetilde{\mathbf{W}}_{d_k,f},
\]
where each $m_{i,k}(s)$ is a complex scalar coefficient of the precoding matrix to be designed.

The wireless channel is enhanced by a passive RIS with $G$ elements. For simplicity, we assume that direct paths between transmitters and receivers are blocked. However, the RIS-assisted interference nulling approach in such a setup can be extended to systems with direct paths, as shown in~\cite{jiang2022interference}. Let $\bm{h}_{T,j}$ and $\bm{h}_{R,k}$ represent the channel vectors (with dimension $G \times 1$) from the $j$-th transmitter to the RIS and from the RIS to the $k$-th receiver, respectively. These channel coefficients are assumed to be i.i.d. with some continuous distribution (e.g., circularly symmetric Gaussian distribution), and full channel state information (CSI) is available to both the server and users.
The RIS is configured by the phase-shift vector 
\[
    \bm{v} = [e^{j\omega_1}, e^{j\omega_2}, \dots, e^{j\omega_G}]^T \in \mathbb{C}^G,
\]
where $\omega_i \in [0, 2\pi)$ represents the phase-shift coefficient of the $i$-th RIS element. The vector $\bm{v}$ can be reconfigured in each time slot. Each user $k$ receives a signal reflected by the RIS during time slot $s$, expressed as:
\begin{subequations}
\begin{align}
    Y_k(s)&=\sum_{j=1}^L \bm{h}_{R,k}^T \operatorname{diag}(\bm{v}) \bm{h}_{T,j} X_j(s) + n_k(s)
    \label{receive signal Y_k}\\
    &=\sum_{j=1}^L \bm{a}_{k,j}^T \bm{v} X_j(s) + n_k(s),
    \label{eq:receive signal Y_k(cascaded)}
\end{align}
\end{subequations}
where $\bm{a}_{k,j} \triangleq \operatorname{diag}(\bm{h}_{T,j}) \bm{h}_{R,k}$, and $n_k(s)$ is the additive white Gaussian noise at receiver $k$.

In this work, we investigate a RIS-assisted multi-antenna coded caching system and aim to characterize and maximize the \emph{sum Degrees of Freedom (sum-DoF)} under linear one-shot delivery. The sum-DoF captures the average number of users that can be served simultaneously per transmission slot. In the high-SNR regime, each coded packet, transmitted at rate $\log P + o(\log P)$ using Gaussian signaling, effectively conveys one DoF. Our goal is to design delivery strategies that maximize the achievable sum-DoF for all possible user demand patterns, while maintaining a low subpacketization level to ensure practical implementability.

\subsection{MAPDA}
For the multi-antenna coded caching system (the same system as in this paper but without RIS), a combinatorial structure, MAPDA, was proposed in~\cite{yang2023multiple,namboodiri2023extended}, 
to construct coded caching schemes with uncoded cache placement and one-shot ZF-based linear delivery.

\begin{definition}[MAPDA, \cite{yang2023multiple,namboodiri2023extended}]
\label{defini:MAPDA}
 For any positive integers $L$, $K$, $F$, $Z$, and $S$, a $F \times K$ array $\mathbf{P} = (\mathbf{P}(j, k))_{j \in [F] , k \in [K]}$ composed of ``$*$'' and $[S]$ is called a $(L, K, F, Z, S)$ multiple-antenna placement delivery array (MAPDA) if it satisfies the following conditions: 
\begin{itemize}
    \item[C1:] The symbol $*$ appears $Z$ times in each column;
    \item[C2:] Each integer $s \in [S]$ occurs at least once in the array;
    \item[C3:] Each integer $s$ appears at most once in each column;
    \item[C4:] For any integer $s \in [S]$, define $\mathbf{P}^{(s)}$ to be the subarray of $\mathbf{P}$ including the rows and columns containing $s$, and let $r'_s \times r_s$ denote the dimensions of $\mathbf{P}^{(s)}$. The number of integer entries in each row of $\mathbf{P}^{(s)}$ is less than or equal to $\min\{L, K\}$, i.e.,
  {\small\[
    \left|\{k_1 \in [r_s] \mid \mathbf{P}^{(s)}(f_1, k_1) \in [S]\}\right| \leq \min\{L, K\}, \quad \forall f_1 \in [r'_s].
    \]}
    %%改成至多r个子阵列
\end{itemize}

If each integer appears $g$ times in MAPDA $\mathbf{P}$, then $\mathbf{P}$ is a $g$-regular MAPDA, denoted by $g$-$(L, K, F, Z, S)$ MAPDA. Given a $g$-MAPDA, we can derive a multi-antenna coded caching scheme for a system consisting of a server with $L$ antennas and $K$ single-antenna users, where the cache-to-library ratio satisfies $M/N = Z/F$, and the sum-DoF is $g$. Note that the maximum achievable sum-DoF under MAPDA structure is upper bounded by $min\{K,KM/N + L\}$.

\hfill $\square$ 
\end{definition} 
By definition, each MAPDA  consists of ``$*$'' symbols and integers, where each column represents a user and each row corresponds to a packet of a file. A ``$*$'' in position $(i,j)$ indicates that user $j$ stores the $i$-th packet of each file. Each integer in the array represents a multicast message transmitted during a single time slot.  Sometimes, for ease of notation, we can also represent each entry  in the MAPDA as a vector or a set. 

In addition, each MAPDA corresponds to a specific coded caching scheme where the cache ratio $ \frac{M}{N} $ equals $ \frac{Z}{F} $. This process is summarized in Algorithm \ref{algorithm:MAPDA}.

\begin{algorithm}
\caption{Caching Scheme Based on MAPDA in \cite{yang2023multiple,namboodiri2023extended}}
\begin{algorithmic}[1]
\Procedure{Placement}{P, $\Wc$}
    \State Split each file $W_n \in W$ into $F$ packets, i.e., $W_n = \{W_{n,f} | f = 1, 2, \dots, F\}$.
    \For{$k \in [K]$}
        \State $Z_k \leftarrow \{W_{n,f} | \text{P}(f, k) = \ast, n \in [N], f \in [F]\}$.
    \EndFor
\EndProcedure
\Procedure{Delivery}{P, $\Wc, d$}
    \For{$s = 1, 2, \dots, S$}
    \State Server uses $L$ antennas to send $\mathbf{W}_{{d_k},j}$ where P$(j,k)=s$ to the users.
    \EndFor
\EndProcedure
\end{algorithmic}
\label{algorithm:MAPDA}
\end{algorithm}

Note that when $L = 1$, the MAPDA reduces to the PDA presented in~\cite{yan2017placement}, which was originally proposed for centralized coded caching over shared-link SISO systems. In a PDA, $r'_s = r_s$.  Note that  when $L=1$,  for each   integer in   $[S]$,  the minimum subarrary of a PDA containing all entries equal to this integer is called a clique. 

We now review some PDAs and MAPDAs that will be used later.

\begin{lem}[Maximum DoF~\cite{yang2023multiple}]
\label{lem:MaxDoF}
 Under the MAPDA structure, the maximum achievable sum-DoF  is $\min\{K, L + K M/N \}$, when $KM/N$ is an integer. 
\end{lem}

\begin{lem}[MN PDA\cite{maddah2014fundamental}]
\label{lem:MNPDA}
    For any positive integers $ K $ and $ t $ with $ t < K $, there exists a $(K, \binom{K}{t}, \binom{K-1}{t-1}, \binom{K}{t+1} )$ PDA with the maximum DoF $t+1$.
\end{lem}  

\begin{construction}[MN PDA~\cite{maddah2014fundamental}]
\label{const:MN PDA}
The MN PDA $ \mathbf{P} = (\mathbf{P}(\mathcal{T}, k))_{\mathcal{T} \in \binom{[K]}{t} , k \in [K]}$ achieving the sum-DoF in Lemma \ref{lem:MNPDA} can be constructed as\footnote{ Notice that, for the sake of ease notation, sometimes we
also express the non-star entries in a PDA by sets or vectors
rather than integers.}
\begin{align}
\label{eq-MN-constr}
    \mathbf{P}(\Tc, k) = 
\begin{cases} 
\ast & \text{if } k \in \Tc, \\
\Tc \cup \{k\} & \text{otherwise}.
\end{cases}
\end{align}
\end{construction}

\begin{lem}[SMK MAPDA\cite{shariatpanahi2016multi}]
\label{lem:SMK MAPDA}
For any positive integers $L_1,~ K $ and $ t $, there exists a 
$(L_1, K, \binom{K}{t} \binom{K-t-1}{L_1-1},\\ \binom{K-1}{t-1} \binom{K-t-1}{L_1-1}, \binom{K}{t+L_1} \binom{t+L_1-1}{t})$
MAPDA with the maximum DoF $t+L_1$.
\end{lem}

\begin{construction}[SMK MAPDA~\cite{shariatpanahi2016multi}]
The SMK MAPDA $\mathbf{P} = (\mathbf{P}(\mathcal{T},\mathcal{L}), k))$ where $\mathcal{T}\in{\binom{[K]}{t}},\mathcal{L}\in \binom{[K-t-1]}{L_1-1}, k\in[K]$, achieving the DoFs of Lemma \ref{lem:SMK MAPDA} can be constructed as 
\begin{align}
\label{eq:SMK MAPDA}
        \mathbf{P}((\mathcal{T},\mathcal{L}), k) = 
        \begin{cases} 
	\ast & \text{if } k \in \Tc, \\
	(\mathcal{S}(\mathcal{T},\mathcal{L},k), \text{order}(\mathcal{S}(\mathcal{T},\mathcal{L},k))) & \text{otherwise},
    \end{cases}
\end{align}
where $\mathcal{S}(\mathcal{T},\mathcal{L},k)=\Tc \cup ([K]\setminus(\mathcal{T}\cup\{k\})[\mathcal{L}])\cup\{k\}$ and  $ \text{order}(\mathcal{S}) $ as the order of appearance of the set $\mathcal{S}$ in each column.
\label{const:SMK MAPDA}
\end{construction}

\begin{ex}
For $K=7$ and $t=1$, we have a $(7,7,1,21)$ MN PDA in Table \ref{tab:MN PDA example}. 
\begin{table}[ht]
\centering
\[
\begin{array}{|c|c|c|c|c|c|c|c|}
\hline
 & 1 & 2 & 3 & 4 & 5 & 6 & 7 \\ 
\hline
1 & * & \{1,2\} & \{1,3\} & \{1,4\} & \{1,5\} & \{1,6\} & \{1,7\} \\ 
\hline
2 & \{1,2\} & * & \{2,3\} & \{2,4\} & \{2,5\} & \{2,6\} & \{2,7\} \\ 
\hline
3 & \{1,3\} & \{2,3\} & * & \{3,4\} & \{3,5\} & \{3,6\} & \{3,7\} \\ 
\hline
4 & \{1,4\} & \{2,4\} & \{3,4\} & * & \{4,5\} & \{4,6\} & \{4,7\} \\ 
\hline
5 & \{1,5\} & \{2,5\} & \{3,5\} & \{4,5\} & * & \{5,6\} & \{5,7\} \\ 
\hline
6 & \{1,6\} & \{2,6\} & \{3,6\} & \{4,6\} & \{5,6\} & * & \{6,7\} \\ 
\hline
7 & \{1,7\} & \{2,7\} & \{3,7\} & \{4,7\} & \{5,7\} & \{6,7\} & * \\ 
\hline
\end{array}
\]
\caption{A $(7,7,1,21)$ MN PDA}
\label{tab:MN PDA example}
\end{table}
\label{ex:MN PDA}
\end{ex}

\begin{ex}
    For $K=7,t=1,L_1=2$,we have a $(2,7,35,5,70)$ SMK MAPDA in Table~\ref{tab:SMK MAPDA example}.
    \begin{table}[ht]
\vspace{0.1in}
\centering
\hspace{-0.3cm}
\resizebox{0.6\textwidth}{!}{
\begin{tabular}{|c|c|c|c|c|c|c|c|}
\hline
 & 1 & 2 & 3 & 4 & 5 & 6 & 7 \\ 
\hline
1 & * & (\{1,2,3\},1) & (\{1,2,3\},1) & (\{1,2,4\},1) & (\{1,2,5\},1) & (\{1,2,6\},1) & (\{1,2,7\},1) \\ 
\hline
2 & (\{1,2,3\},1) & * & (\{1,2,3\},2) & (\{1,2,4\},2) & (\{1,2,5\},2) & (\{1,2,6\},2) & (\{1,2,7\},2) \\ 
\hline
3 & (\{1,2,3\},2) & (\{1,2,3\},2) & * & (\{1,3,4\},1) & (\{1,3,5\},1) & (\{1,3,6\},1) & (\{1,3,7\},1) \\ 
\hline
4 & (\{1,2,4\},1) & (\{1,2,4\},1) & (\{1,3,4\},1) & * & (\{1,4,5\},1) & (\{1,4,6\},1) & (\{1,4,7\},1) \\ 
\hline
5 & (\{1,2,5\},1) & (\{1,2,5\},1) & (\{1,3,5\},1) & (\{1,4,5\},1) & * & (\{1,5,6\},1) & (\{1,5,7\},1) \\ 
\hline
6 & (\{1,2,6\},1) & (\{1,2,6\},1) & (\{1,3,6\},1) & (\{1,4,6\},1) & (\{1,5,6\},1) & * & (\{1,6,7\},1) \\ 
\hline
7 & (\{1,2,7\},1) & (\{1,2,7\},1) & (\{1,3,7\},1) & (\{1,4,7\},1) & (\{1,5,7\},1) & (\{1,6,7\},1) & * \\ 
\hline
1 & * & (\{1,2,4\},2) & (\{1,3,4\},2) & (\{1,3,4\},2) & (\{1,3,5\},2) & (\{1,3,6\},2) & (\{1,3,7\},2) \\
\hline
2 & (\{1,2,4\},2) & * & (\{2,3,4\},1) & (\{2,3,4\},1) & (\{2,3,5\},1) & (\{2,3,6\},1) & (\{2,3,7\},1) \\
\hline
3 & (\{1,3,4\},1) & (\{2,3,4\},1) & * & (\{2,3,4\},2) & (\{2,3,5\},2) & (\{2,3,6\},2) & (\{2,3,7\},2) \\
\hline
4 & (\{1,3,4\},2) & (\{2,3,4\},2) & (\{2,3,4\},2) & * & (\{2,4,5\},1) & (\{2,4,6\},1) & (\{2,4,7\},1) \\
\hline
5 & (\{1,3,5\},1) & (\{2,3,5\},1) & (\{2,3,5\},1) & (\{2,4,5\},1) & * & (\{2,5,6\},1) & (\{2,5,7\},1) \\
\hline
6 & (\{1,3,6\},1) & (\{2,3,6\},1) & (\{2,3,6\},1) & (\{2,4,6\},1) & (\{2,5,6\},1) & * & (\{2,6,7\},1) \\
\hline
7 & (\{1,3,7\},1) & (\{2,3,7\},1) & (\{2,3,7\},1) & (\{2,4,7\},1) & (\{2,5,7\},1) & (\{2,6,7\},1) & * \\
\hline
1 & * & (\{1,2,5\},2) & (\{1,3,5\},2) & (\{1,4,5\},2) & (\{1,4,5\},2) & (\{1,4,6\},2) & (\{1,4,7\},2) \\
\hline
2 & (\{1,2,5\},2) & * & (\{2,3,5\},2) & (\{2,4,5\},2) & (\{2,4,5\},2) & (\{2,4,6\},2) & (\{2,4,7\},2) \\
\hline
3 & (\{1,3,5\},2) & (\{2,3,5\},2) & * & (\{3,4,5\},1) & (\{3,4,5\},1) & (\{3,4,6\},1) & (\{3,4,7\},1) \\
\hline
4 & (\{1,4,5\},1) & (\{2,4,5\},1) & (\{3,4,5\},1) & * & (\{3,4,5\},2) & (\{3,4,6\},2) & (\{3,4,7\},2) \\
\hline
5 & (\{1,4,5\},2) & (\{2,4,5\},2) & (\{3,4,5\},2) & (\{3,4,5\},2) & * & (\{3,5,6\},1) & (\{3,5,7\},1) \\
\hline
6 & (\{1,4,6\},1) & (\{2,4,6\},1) & (\{3,4,6\},1) & (\{3,4,6\},1) & (\{3,5,6\},1) & * & (\{3,6,7\},1) \\
\hline
7 & (\{1,4,7\},1) & (\{2,4,7\},1) & (\{3,4,7\},1) & (\{3,4,7\},1) & (\{3,5,7\},1) & (\{3,6,7\},1) & * \\
\hline
1 & * & (\{1,2,6\},2) & (\{1,3,6\},2) & (\{1,4,6\},2) & (\{1,5,6\},2) & (\{1,5,6\},2) & (\{1,5,7\},2) \\
\hline
2 & (\{1,2,6\},2) & * & (\{2,3,6\},2) & (\{2,4,6\},2) & (\{2,5,6\},2) & (\{2,5,6\},2) & (\{2,5,7\},2) \\
\hline
3 & (\{1,3,6\},2) & (\{2,3,6\},2) & * & (\{3,4,6\},2) & (\{3,5,6\},2) & (\{3,5,6\},2) & (\{3,5,7\},2) \\
\hline
4 & (\{1,4,6\},2) & (\{2,4,6\},2) & (\{3,4,6\},2) & * & (\{4,5,6\},1) & (\{4,5,6\},1) & (\{4,5,7\},1) \\
\hline
5 & (\{1,5,6\},1) & (\{2,5,6\},1) & (\{3,5,6\},1) & (\{4,5,6\},1) & * & (\{4,5,6\},2) & (\{4,5,7\},2) \\
\hline
6 & (\{1,5,6\},2) & (\{2,5,6\},2) & (\{3,5,6\},2) & (\{4,5,6\},2) & (\{4,5,6\},2) & * & (\{4,6,7\},1) \\
\hline
7 & (\{1,5,7\},1) & (\{2,5,7\},1) & (\{3,5,7\},1) & (\{4,5,7\},1) & (\{4,5,7\},1) & (\{4,6,7\},1) & * \\
\hline
1 & * & (\{1,2,7\},2) & (\{1,3,7\},2) & (\{1,4,7\},2) & (\{1,5,7\},2) & (\{1,6,7\},2) & (\{1,6,7\},2) \\
\hline
2 & (\{1,2,7\},2) & * & (\{2,3,7\},2) & (\{2,4,7\},2) & (\{2,5,7\},2) & (\{2,6,7\},2) & (\{2,6,7\},2) \\
\hline
3 & (\{1,3,7\},2) & (\{2,3,7\},2) & * & (\{3,4,7\},2) & (\{3,5,7\},2) & (\{3,6,7\},2) & (\{3,6,7\},2) \\
\hline
4 & (\{1,4,7\},2) & (\{2,4,7\},2) & (\{3,4,7\},2) & * & (\{4,5,7\},2) & (\{4,6,7\},2) & (\{4,6,7\},2) \\
\hline
5 & (\{1,5,7\},2) & (\{2,5,7\},2) & (\{3,5,7\},2) & (\{4,5,7\},2) & * & (\{5,6,7\},1) & (\{5,6,7\},1) \\
\hline
6 & (\{1,6,7\},1) & (\{2,6,7\},1) & (\{3,6,7\},1) & (\{4,6,7\},1) & (\{5,6,7\},1) & * & (\{5,6,7\},2) \\
\hline
7 & (\{1,6,7\},2) & (\{2,6,7\},2) & (\{3,6,7\},2) & (\{4,6,7\},2) & (\{5,6,7\},2) & (\{5,6,7\},2) & * \\
\hline
\end{tabular}
}
\caption{A $(2,7,35,5,70)$ SMK MAPDA}
\label{tab:SMK MAPDA example}
\end{table}
\label{ex:SMK MAPDA example}
\end{ex}

 \section{Tangent-Space Enhanced RIS-Assisted Interference Nulling algorithm}
% \section{Improved RIS-assisted Interference Nulling Algorithm}
\label{sec:nulling alg}

In this section, we present an enhanced interference nulling algorithm for RIS-assisted multi-antenna coded caching systems. The objective is to efficiently compute the phase-shift vector at the RIS that suppresses designated interference paths from transmit antennas to users, while adhering to the unit-modulus constraint imposed by passive RIS hardware. Leveraging recent advances in Riemannian optimization, we incorporate tangent-space projection into a refined alternating projection framework to accelerate convergence. We begin by formulating the interference nulling problem from a geometric perspective and then describe  the proposed algorithm. Numerical results demonstrate that the proposed method  outperforms existing approaches in terms of convergence speed.

\subsection{Problem Formulation and Geometric Interpretation}
\label{sub:opt problem}
% \subsection{Incorporating Tangent Space Projection in RIS Nulling}
In this section, we consider the problem of designing the passive RIS phase-shift vector $\bm{v} \in \mathbb{C}^{G}$ (where the RIS contains $G$ elements) 
to eliminate undesired interference paths in a RIS-assisted $K$-user MISO system. To suppress the interference between the $j$-th transmit antenna and user $k$, the received signal must satisfy the constraint $\bm{a}_{k,j}^T \bm{v} = 0$, where $\bm{a}_{k,j}= \operatorname{diag}(\bm{h}_{T,j}) \bm{h}_{R,k} \in \mathbb{C}^{G}$ 
denotes the cascaded channel from antenna $j$ to user $k$ via the RIS. Let $\bm{A} \in \mathbb{C}^{G \times p}$ be the interference channel matrix, whose columns are the cascaded channel vectors $\bm{a}_{k,j}$ corresponding to all $p$ interference paths to be nullified. More precisely, for user $k$, we assume that the interference paths to be eliminated originate from the transmit antennas $j_{k,1}, j_{k,2}, \dots, j_{k,q_k}$, where $j_{k,q} \in [L]$. Here $q_k$ denotes the number of paths to be nullified for user $k$. Considering all users, the total number of interference paths to be eliminated is $\sum_{k \in [K]} q_k=p$.
Denote 
$   \bm{A} = [\bm{a}_{1,j_{1,1}}, \dots, \bm{a}_{1,j_{1,q_1}}, \bm{a}_{2,j_{2,1}}, \dots, \bm{a}_{K,j_{K,q_K}}], $ 
where each column of $\bm{A}$ represents an interference path that needs to be nullified. 
As shown in~\cite{jiang2022interference}, nullifying $p$ interference paths typically requires the number of RIS elements $G$ to be slightly greater than $2p$. For simplicity, we adopt the approximation $G = 2p$ in our analysis. 
\iffalse 
The interference nulling objective can then be formulated as finding a solution that satisfies  
\begin{align}
    \Sc_1 = \{\bm{A}^T \bm{v} = 0\},
\end{align}  
ensuring that the interference paths are completely suppressed.

Additionally, since the RIS is passive, the phase-shift vector $\bm{v}$ must satisfy the unit modulus constraint:  
\begin{align}
    \Sc_2 = \{|v_i| = 1\}, \quad \forall i.
\end{align}  
\fi 

The interference nulling problem can then be formulated as the following feasibility problem:
\begin{align}
      \text{Find } \bm{v} \in \mathbb{C}^G \text{ such that } 
    \begin{cases}
        (i). \  \bm{A}^T \bm{v} = 0 .\\
       (ii). \  |v_i| = 1,\quad \forall i \in [G].
    \end{cases}
    \label{eq:inteference nulling problem}
\end{align}
where the first constraint enforces interference suppression, and the second imposes the unit-modulus constraint intrinsic to passive RIS elements.

The optimization problems of this type, which aim to find structured solutions under non-convex constraints, frequently appear in compressed sensing and low-rank matrix recovery. In these domains, iterative hard thresholding (IHT) methods~\cite{blanchard2014conjugate,blanchard2015cgiht,kyrillidis2014matrix,tanner2013normalized} have been widely studied as simple yet effective projection-based algorithms for recovering sparse or low-rank signals. However, classical IHT variants, such as normalized IHT (NIHT), often exhibit slow convergence due to their reliance on global gradient directions, which neglect the underlying geometric structure of the feasible region.

To address this limitation, recent work has explored Riemannian optimization techniques~\cite{wei2016guarantees,boumal2023introduction}, wherein the search direction is confined to the tangent space of the constraint manifold at the current iterate. This approach forms the basis of the Riemannian Gradient Descent (RGrad) algorithm, which leverages the local geometry of the feasible set to achieve faster convergence and improved search directionality. 
Inspired by the core principle of RGrad, we propose an enhanced alternating projection algorithm for RIS-assisted interference nulling. In particular, the update direction is geometrically refined by eliminating the component orthogonal to the linear nulling subspace, thereby approximating a tangent-space projection. This geometry-aware correction enables the algorithm to take larger and more effective update steps than the existing alternating projection method in~\cite{jiang2022interference}, resulting in faster convergence. The detailed design and analysis of the proposed algorithm are presented in the remainder of this section.

\subsection{Tangent-Space Enhanced Alternating Projection algorithm}
To efficiently solve the interference nulling problem described in Section~\ref{sub:opt problem}, we propose a tangent-space enhanced alternating projection algorithm. Specifically, given the interference channel matrix $\bm{A} \in \mathbb{C}^{G \times p}$ and the feasible sets
\[
    \Sc_1 = \{\bm{v} \in \mathbb{C}^G : \bm{A}^T \bm{v} = 0\}; \quad
    \Sc_2 = \{\bm{v} \in \mathbb{C}^G : |v_i| = 1,~ \forall i\in [G]\},
\]
our goal is to find a vector $\bm{v}$ that lies in the intersection $\Sc_1 \cap \Sc_2$. 
To this end, we design an alternating projection method that iteratively refines $\bm{v}$ via projections onto $\Sc_1$ and $\Sc_2$, while incorporating tangent-space geometry to accelerate convergence. Below, we present the algorithmic details.

The algorithm proposed in~\cite{jiang2022interference} solves this problem by iteratively projecting $\bm{v}$ onto the sets $\Sc_1$ and $\Sc_2$. Specifically, the projection onto $\Sc_1$ is given by:  
\begin{align}
    \Pi_{\Sc_1}(\bm{v}) = \bm{v} - \bm{A}^* (\bm{A}^T \bm{A}^*)^{-1} \bm{A}^T \bm{v},
\label{projection_to_Sc_1}
\end{align}  
which ensures that the interference constraints are satisfied. The projection onto $\Sc_2$ is defined as:  
\begin{align}
    \Pi_{\Sc_2}(\bm{v}) = \frac{\bm{v}}{|\bm{v}|},
\label{projection_to_Sc_2}
\end{align}  
which enforces the unit modulus constraint on each element of $\bm{v}$. The algorithm alternates between these two projections, gradually converging to a solution that lies at the intersection of $\Sc_1$ and $\Sc_2$.

Unlike the classical alternating projection method in~\cite{jiang2022interference}, the proposed approach leverages tangent space projection to enable larger and more effective update steps. This geometry-aware strategy significantly accelerates convergence in RIS-assisted interference nulling scenarios. The pseudocode of the proposed scheme is given in Algorithm~\ref{alg:Improved_Algorithm}.\footnote{\label{foot:extend to LOS}The algorithm is naturally extensible to scenarios where direct transmission paths exist, with some minor adjustments to the projection formulation.}

\begin{algorithm}
\caption{Tangent-Space Enhanced Alternating Projection Algorithm}
\label{alg:Improved_Algorithm}  % Adjusted to use an underscore for consistency
\begin{algorithmic}[1] % Enable numbering
\State Initialize reflection coefficients $\bm{v}$, channel matrix $\bm{A}$, and number of iterations $m$. 
\State Start with initial value $\bm{v} = \bm{v}^0 \in \mathcal{S}_2$. 
\State \textbf{while} $t \leq m$ \textbf{and} the interference has not been nullified \textbf{do}
    \Statex $ \bm{y}^t = \bm{v}^t - \Pi_{\mathcal{S}_1}(\bm{v}^t) $;
    \Statex $ \bm{m}^t = \bm{y}^t - \Pi_{\bm{v}^t}(\bm{y}^t) $;
    \Statex $ \tilde{\bm{v}}^t = \bm{v}^t - 2\bm{m}^t $;
    \Statex $ \bm{v}^{t+1} = \Pi_{\mathcal{S}_2}(\tilde{\bm{v}}^t) $.
% \State \textbf{end while}
\State \textbf{if} iteration limit reached or interference nullified \textbf{then}
    \Statex Stop iteration. 
\State \textbf{end if}
\State Output the resulting reflection coefficients $\bm{v}^{t+1}$. 
\end{algorithmic}
\end{algorithm}

The proposed method in Algorithm~\ref{alg:Improved_Algorithm} enhances the classical alternating projection technique by incorporating a geometry-aware correction mechanism that improves convergence in RIS-assisted interference nulling problems. 
Drawing inspiration from Riemannian optimization, the proposed algorithm refines the update direction by incorporating projections onto the tangent space of $\mathcal{S}_2$ prior to enforcing the modulus constraint. At each iteration $t$, the orthogonal residual between the current estimate $\boldsymbol{v}^t$ and the subspace $\mathcal{S}_1$ is computed as:
\begin{align*}
    \bm{y}^t &= \bm{v}^t - \Pi_{\mathcal{S}_1}(\bm{v}^t) = \bm{A}^* (\bm{A}^T \bm{A}^*)^{-1} \bm{A}^T \bm{v}^t.
\end{align*}
The component of $\bm{y}^t$ orthogonal to $\boldsymbol{v}^t$ is then extracted as:
\begin{align*}
    \bm{m}^t &=\bm{y}^t - \Pi_{\bm{v}^t}(\bm{y}^t)= \bm{y}^t - \operatorname{Re}[(\bm{v}^t)^* \odot \bm{y}^t] \odot \bm{v}^t.
\end{align*}
The mapping $\Pi_{\bm{v}^t}(\bm{y}^t)$ represents the orthogonal projection of the residual vector $\bm{y}^t$ onto the direction of the current iterate $\bm{v}^t$. This operation extracts the real-valued component of $\bm{y}^t$ aligned with $\bm{v}^t$, thereby eliminating contributions that are orthogonal to this direction. Such decomposition plays a critical role in steering updates along the tangent space of the constraint manifold, thus enhancing geometric consistency and accelerating convergence.

Formally, the projection is defined as
\begin{equation*}
\Pi_{\bm{v}^t}(\bm{y}^t) = \operatorname{Re} \left[ (\bm{v}^t)^* \odot \bm{y}^t \right] \odot \bm{v}^t,
\end{equation*}
where $\odot$ denotes element-wise multiplication.

By isolating the correctable component of the residual, this mapping enables more focused updates, thereby improving both the efficiency and stability of the alternating projection algorithm.

The update step is performed by moving along the negative direction of $\boldsymbol{m}^t$, scaled by a factor of $2$, followed by projection back onto $\mathcal{S}_2$:
\begin{align*}
    \tilde{\bm{v}}^t &= \bm{v}^t - 2\bm{m}^t, \\
    \bm{v}^{t+1} &= \frac{\tilde{\bm{v}}^t}{|\tilde{\bm{v}}^t|}.
\end{align*}
Here, the factor $2$ is introduced to enlarge the step size along the tangent direction, thereby enabling faster traversal toward the feasible region and promoting accelerated convergence. While larger scaling factors can help escape regions of slow improvement, they may also introduce instability if they are overly aggressive.

\subsection{Initialization Strategy}
For a fair and consistent comparison with the reference algorithm in~\cite{jiang2022interference}, we adopt the same initialization strategy for the phase shift vector as described in their work. Specifically, rather than employing a fully random unit-modulus initialization, we use the eigenvector-based method proposed in~\cite{jiang2022interference}, where the initial vector is chosen as the principal eigenvector of the aggregated signal covariance matrix formed from the effective channel vectors. Prior results have demonstrated that this initialization significantly accelerates convergence and enhances performance compared to random initialization~\cite{jiang2022interference}. Accordingly, we employ this initialization scheme for both algorithms in our simulations to ensure consistency and a fair basis for comparison.

\subsection{Convergence Analysis}
\begin{thm}[Local Convergence of the Tangent-Space Enhanced AP Algorithm]
Let $\mathcal{S}_1 = \{\bm{v} \in \mathbb{C}^G : \bm{A}^\top \bm{v} = 0\}$ and $\mathcal{S}_2 = \{\bm{v} \in \mathbb{C}^G : |v_i| = 1,~ \forall i \in [G]\}$.
Suppose their intersection $\mathcal{S} := \mathcal{S}_1 \cap \mathcal{S}_2$ is nonempty, and let $\bm{v}^\star \in \mathcal{S}$ be a feasible point.
Then, for any initial point $\bm{v}^0$ sufficiently close to $\bm{v}^\star$, the sequence $\{\bm{v}^t\}$ generated by the tangent-space enhanced alternating projection (TS--AP) iteration 
converges locally to a fixed point $\bm{v}^\infty \in \mathcal{S}$.
\end{thm}

\begin{proof}
Throughout this section, we regard the unit-modulus constraint
\[
\mathcal M := \mathcal S_2
 =\bigl\{\bm v\in\mathbb C^{G}\;:\;|v_i|=1,\;i=1,\dots,G\bigr\}
\]
as a smooth (real) $G$-dimensional Riemannian manifold
(endowed with the canonical metric inherited from~$\mathbb C^{G}$).
The violation of the linear constraint
$\mathcal S_1 = \ker(\bm{A}^{\!\top})$
is measured by the smooth cost
\[
f(\bm v) = \tfrac{1}{2}\left\|\bm{A}^{\!\top}\bm v\right\|_2^{2},
\qquad
\bm v \in \mathcal M .
\]

We verify the assumptions in Proposition 4.6 of \cite{boumal2023introduction}:

\textbf{A1 (lower boundedness).}  
The objective function is non-negative for all $\bm{v} \in \mathcal{S}_2$:
\[
f(\bm v) = \tfrac{1}{2}\left\|\bm{A}^{\!\top}\bm v\right\|_2^{2} \ge 0.
\]

\textbf{A2 (sufficient decrease).}  
The TS--AP algorithm updates $\bm{v}^{t+1}$ as follows:
\[
\bm{v}^{t+1} = \Pi_{\mathcal{S}_2} \left( \bm{v}^t - 2\bm{m}^t \right),
\quad \text{where } \bm{m}^t := \bm{y}^t - \Pi_{\bm{v}^t}(\bm{y}^t),\;
\bm{y}^t := \bm{v}^t - \Pi_{\mathcal{S}_1}(\bm{v}^t).
\]
Here, $\Pi_{\bm{v}^t}(\bm{y}^t)$ denotes the projection of $\bm{y}^t$ onto the direction of $\bm{v}^t$ (i.e., the normal space), and $\bm{m}^t$ is orthogonal to $\bm{v}^t$, thus lying in the tangent space $T_{\bm{v}^t}\mathcal{M}$.

This constitutes a Riemannian descent direction on the manifold $\mathcal{M}$.  
Under mild smoothness assumptions and non-expansiveness of projections, there exists a constant $c > 0$ such that:
\[
f(\bm{v}^{t}) - f(\bm{v}^{t+1}) \ge c \cdot \|\mathrm{grad} f(\bm{v}^t)\|^2.
\]

By Proposition 4.6 in \cite{boumal2023introduction}, it follows that
\[
\lim_{t \to \infty} \|\mathrm{grad} f(\bm{v}^t)\| = 0.
\]
\end{proof}

\subsection{Simulation Results and Analysis}
In this section, we compare the proposed Algorithm~\ref{alg:Improved_Algorithm} with the RIS-assisted interference nulling scheme introduced in~\cite{jiang2022interference}. We consider the system model illustrated in Fig.~\ref{fig:system model}, where user-side caching is temporarily omitted, and the number of transmitters is equal to the number of users, i.e., \(K = L\). This setup corresponds to a \(K\)-transmitter, \(K\)-user interference channel, where the RIS is utilized to cancel \(K - 1\) interference paths for each user, resulting in a total of \(K(K - 1)\) interference links to be nullified. In our experiments, we set \(K = 10\) and configure the RIS to consist of \(G = 300\) reflecting elements.

The channel coefficients are assumed to be independent and identically distributed (i.i.d.) according to the standard complex Gaussian distribution. Specifically, each entry of the transmitter-side and receiver-side channel vectors, denoted by \(\bm{h}_{T,j}\) and \(\bm{h}_{R,k}\), respectively, is generated with independent real and imaginary parts drawn from \(\mathcal{N}(0,1)\). Both algorithms are executed for a maximum of 500 iterations. At each iteration, the interference power at user \(k\) is computed as
\[
\sum_{j \neq k} \left|\bm{a}_{k,j} \bm{v}\right|^2,
\]
and the result is reported in decibels (dB). The overall performance metric is the total interference power $\sum_{k \in [K]} \sum_{j \neq k} \left|\bm{a}_{k,j} \bm{v}\right|^2$ across all users.

\begin{figure}
\centering
  \includegraphics[scale=0.5]{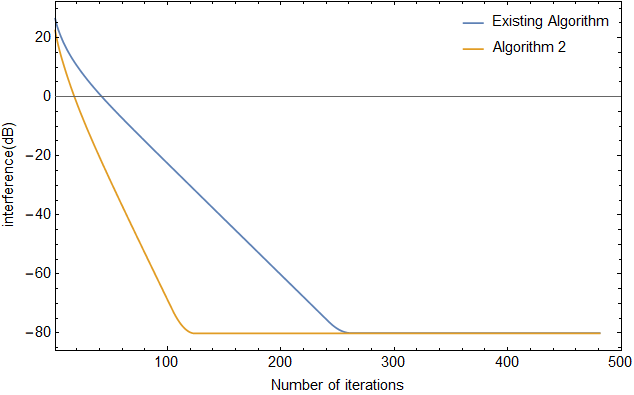}
\caption{Comparison of two algorithms in interference power}
\label{fig:algorithm compare interference}
\end{figure}

As shown in Fig.~\ref{fig:algorithm compare interference}, the proposed Algorithm~\ref{alg:Improved_Algorithm} significantly outperforms the algorithm from~\cite{jiang2022interference}, achieving lower total interference power within the same number of iterations.

% \begin{figure}[htbp]
% \centering
% \centerline{\includegraphics[scale=0.8]{RIS assisted MISO cacing.png}}
% \caption{Example of a figure caption.}
% \label{fig:system model}
% \end{figure}

%\addtolength{\topmargin}{0.01in}

\section{Optimal Grouping Approach}
\label{sec:best group}
\begin{figure}[htbp]
    \centering
    \includegraphics[width=0.8\linewidth]{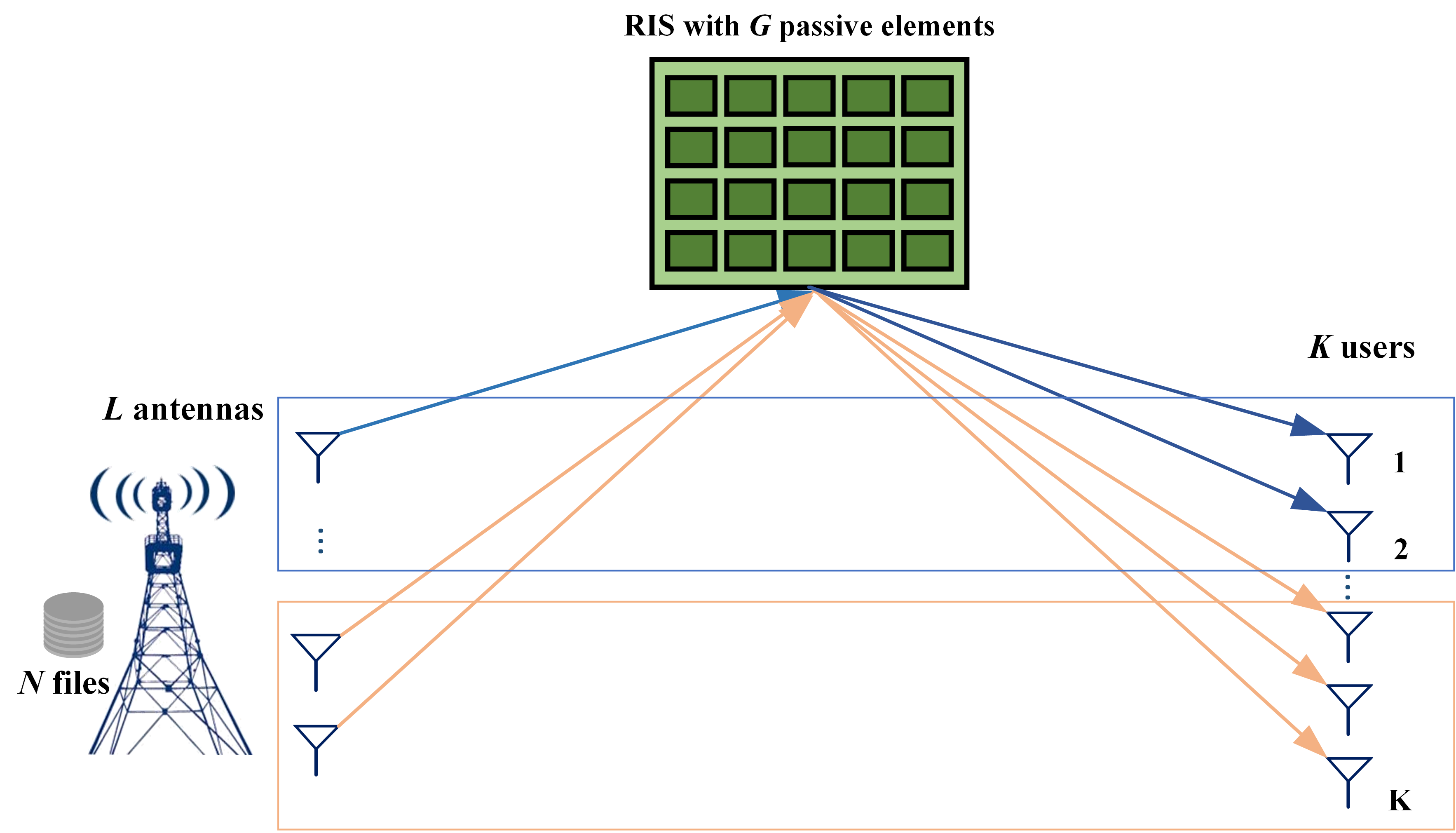}
    \caption{Dividing Users into Interference-Free Groups via RIS. }
    \label{fig:ris_grouping}
\end{figure}

Using the RIS-assisted interference nulling method described in Section~\ref{sec:nulling alg}, the transmissions in each time slot of the delivery phase can be partitioned into multiple interference-free groups. 
Within each group, the designated antennas serve the corresponding users without interference from transmissions in other groups. To illustrate the proposed interference-free grouping strategy, we provide a schematic diagram in Fig.~\ref{fig:ris_grouping}.

This section focuses on determining the optimal grouping strategy that maximizes the sum-DoF, ensuring efficient utilization of system resources.

Consider a system with $L$ total antennas, a cache ratio of $M/N$, and $K$ users. Assume that $L_0 \leq L$ antennas are active during transmission, allowing some antennas to remain ``silent''. These $L_0$ active antennas are further divided into $r$ groups, where each group contains $L_i$ antennas such that $\sum_{i=1}^r L_i = L_0$. Let $t = KM/N$ represent the integer value derived from the cache ratio and the number of users.  

Based on Lemma~\ref{lem:MaxDoF} and the proposed grouping strategy, the sum-DoF is as follows,
% \footnote{\label{foot:big K}For simplicity, we assume that $g$ in~\eqref{eq:dof} does not exceed $K$. If $g > K$, additional virtual users ($g-K$) can be introduced into the system, allowing the same grouping strategy as in the case of $g = K$ to be applied.}  
\begin{align}
    g = \sum_{i=1}^r (L_i + t) = L_0 + tr.
\label{eq:dof}
\end{align}

Given that the number of RIS elements is twice the number of interference paths as explained in Section~\ref{sub:opt problem}, and that the maximum number of users served by each group is $L_i + t$ from Lemma~\ref{lem:MaxDoF}, the required number of RIS elements $G$ must satisfy:  
\begin{align}
    G = 2\sum_{i=1}^r (L_i + t)(L_0 - L_i).
\label{eq:G elements}
\end{align}

One possible formulation on searching the optimal grouping method is to maximize the sum-DoF, given any fixed number of RIS elements. However, given the number of RIS elements, one can determine how many interference paths can be canceled, but not which specific ones. This makes the design problem a complex and unstructured search over all user groupings and interference patterns. 
Hence, in this paper
we consider an alternative optimization problem that is more tractable: given a target sum-DoF $ g $, we aim to minimize the required number of RIS elements. 
Given a target sum-DoF $g$, we search for $(L_0,r)$ to  achieve a DoF of at least $g$, while minimizing $G$; %in~\eqref{eq:Gminimum}.
mathematically, the optimization task is to solve for the minimal value of $ G $ subject to the constraints on $ L_0 $, $ t $, and $ g $, as follows:
\begin{equation}
\begin{aligned}
& \underset{r, ( L_i \in \mathbb{Z}^+ :i\in [r])}{\text{minimize}} & & G \\
& \text{subject to} & & L_0 \leq L, \ r \leq L, \ g  \leq  L_0 + tr.
\end{aligned}
\label{eq:min G}
\end{equation}

 \begin{thm}
 \label{thm optimal group}
     Given the sum-DoF $g=L_0+tr$,  the number of groups $r$, and the number of active antennas  $L_0$, the grouping scheme with the minimum number of RIS elements is $L_1 = L_0-(r-1)$ and $L_2 =  \dots = L_r = 1$,   
while  the minimum number of RIS elements is 
\begin{align}
G = 2(r-1)((t+2)L_0-r).
\label{eq:Gminimum}
\end{align}
 \end{thm}

The proof of Theorem~\ref{thm optimal group} is presented in Appendix~\ref{sec:proof of thm optimal group}. 

According to Theorem~\ref{thm optimal group}, the task reduces to determining $ r $ and $ L_0 $ that minimize~\eqref{eq:Gminimum}.  
A naive method to search for the optimal $ r $ and $ L_0 $ could be a full-grid brute force method enumerating all integer pairs $(L_0,r)$ with $1\le L_0\le L$ and $1\le r\le L$, then filtering those that satisfy $g=L_0+tr$; %and evaluating $G$; 
 this requires $O(L^2)$ checks in the worst case. Instead, we propose an efficient search method based on the following property of the optimization problem in Theorem~\ref{thm optimal group}, whose proof could be found in Appendix~\ref{sec:Design Process of  alg:Optimal Grouping}.
\begin{proposition}[Property of optimal grouping]
    \label{prop:optimal_grouping_properties}
For a given $(g,t,L)$, the optimal grouping parameters $(L_0,r)$ that minimize the number of RIS elements $G(L_0,r)$ under $g = L_0 + tr$ satisfy that
$L_0$ is either $L_0^{\min}$ or $L_0^{\max}$, where 
\[
L_0^{\min}=L_{\min}+(g-L_{\min})\bmod t, \qquad
L_0^{\max}=L_{\max}-(L_{\max}-g)\bmod t.
\] 
\end{proposition}

 Based on the above property, the proposed endpoint-only procedure is summarized in Algorithm~\ref{alg:Optimal Grouping}. 
 Lines~2--4 of Algorithm~\ref{alg:Optimal Grouping}  check whether the target DoF $g$ lies within the  achievable range for at most $L$ antennas. If $g > L(t+1)$, the target DoF is unattainable, whereas if $g < t+1$, the demand can be met without activating multiple antennas. For a feasible $g$, line~5 computes the range $[L_{\min}:L_{\max}]$ of active antennas that can satisfy the DoF $g $. Lines~6--7 then identify the two endpoints $L_0^{\min}$ and $L_0^{\max}$ within this interval that satisfy the target DoF. If no feasible endpoint exists, lines~8--12 increase $g$ to the smallest achievable DoF greater than the current value. If feasible endpoints exist, lines~14--20 evaluate the required number of RIS elements $G(x)$ at these two endpoints only and output the solution $(g, L_0^\star, r^\star, G^\star)$ corresponding to the smaller $G$.

\begin{algorithm}[t]
\caption{Endpoint-Only Optimal Grouping}
\label{alg:Optimal Grouping}
\begin{algorithmic}[1]
\State \textbf{Input:} $L$, $t$, target DoF $g$; optional lower bound $r_{\min}\!\ge\!1$ (default $=1$).
\If{$g > L(t+1)$} 
    \State \textbf{Output:} infeasible; \Return
\EndIf
\State $L_{\min}\!\gets\!\left\lceil g/(t+1)\right\rceil$, $L_{\max}\!\gets\!\min\{L,\,g-t\}$.
% \If{$L_{\min}>L_{\max}$} (empty interval)
%     \State $\delta \gets (L_{\min}-g)\bmod t$;
%     \State $g \gets g+\delta$; 
%     \State \textbf{goto} 2.
% \EndIf
\State $L_0^{\max}\!\gets\! L_{\max}-(L_{\max}-g)\bmod t$;
\State $L_0^{\min}\!\gets\! L_{\min}+(g-L_{\min})\bmod t$.
\If{$L_0^{\min}>L_0^{\max}$} (no feasible point in interval)
    \State $\delta \gets (L_{\min}-g)\bmod t$;
    \State $g \gets g+\delta$; 
    \State \textbf{goto} 2.
\EndIf
\State Build candidates $\mathcal{C}=\varnothing$.
\For{$x \in \{L_0^{\min},\,L_0^{\max}\}$}
    \State $r\gets (g-x)/t$;
    \If{$r\in\mathbb{Z}$ and $r\ge r_{\min}$}
        \State $G(x)\gets 2(r-1)((t+2)x-r)$; 
        \State add $(x,r,G(x))$ to $\mathcal{C}$.
    \EndIf
\EndFor
\State $(L_0^\star,r^\star,G^\star)\gets \arg\min_{(x,r,G)\in\mathcal{C}} G$.
\State \textbf{Output:} $(g, L_0^\star, r^\star, G^\star)$.
\end{algorithmic}
\end{algorithm}

%A naive full-grid brute force enumerates all integer pairs $(L_0,r)$ with $1\le L_0\le L$ and $1\le r\le L$, then filters those satisfying $g=L_0+tr$ and evaluates $G$; this requires $O(L^2)$ checks in the worst case. 
In conclusion, compared to the full-grid brute force method, 
 our endpoint-only method exploits the fact that the global minimizer of $G=2(r-1)((t+2)L_0-r)$ over feasible $(L_0,r)$ must lie at one of the two congruent endpoints $L_0^{\min},L_0^{\max}$, which are available in closed form; hence, only two evaluations of $G$ are needed. 
If the congruence class is empty within the interval, we adjust $g$ by one modulo step (at most $t-1$) and re-apply the same constant-time check. 
Therefore, our method runs in $O(1)$ time with respect to $L$ (or $O(t)$ if $t$ is treated as a variable), strictly improving upon the $O(L^2)$ full-grid brute force.

\section{RIS-Assisted Grouping Coded Caching Scheme}
\label{sec:general construction}
After solving the grouping optimization problem, we proceed to construct a coded caching scheme with the sum-DoF $g = \min(K, L_{\text{opt}} + t r_{\text{opt}})$ based on the optimal values $L_{\text{opt}} = L_0$ and $r_{\text{opt}} = r$. By leveraging the grouping strategy, the transmissions in each time slot are partitioned into multiple independent transmissions, with each transmission corresponding to a single group.

\subsection{RMAPDA}
Inspired by the MAPDA framework, our scheme can also be represented as an array. Similar to MAPDA, the symbol ``$*$'' denotes cached content at the users. Our objective in the delivery phase is to enable simultaneous broadcasting across $r$ groups. The newly proposed structure for the RIS-assisted multi-antenna coded caching problem is termed RMAPDA.

\begin{definition}[RMAPDA]
\label{defini:RMAPDA}
    For any positive integers $L_0$, $K$, $F$, $Z$, $r$, and $S$, a $F \times K$ array $\mathbf{P}$ composed of ``$*$'' and elements from $[S]$ is called a $(L_0, K, F, Z, r, S)$ RIS-assisted multiple-antenna placement delivery array (RMAPDA) if it satisfies the following conditions:
\begin{itemize}[itemsep=0.1em, parsep=0.1em]
    \item[C1:] The symbol ``$*$'' appears exactly $Z$ times in each column;
    \item[C2:] Each integer from $1$ to $S$ appears at least once in the array;
    \item[C3:] Each integer $s$ appears at most once in each column;
    \item[C4:] For any integer $s \in [S]$, define $\mathbf{P}^{(s)}_1, \mathbf{P}^{(s)}_2, \dots, \mathbf{P}^{(s)}_r$ as the subarrays of $\mathbf{P}$ containing the rows and columns where $s$ appears. These subarrays are mutually disjoint in columns. At most one of these subarrays has integer entries in each row that do not exceed $L_0 - r + 1$, while the remaining subarrays have integer entries in each row of at most 1.
\end{itemize}
\end{definition}
 Conditions C1--C3 are inherited from the definition of MAPDA. In addition to these, the proposed scheme must satisfy an additional structural constraint that reflects both the physical limitations of RIS-assisted multi-antenna systems and the design principles underlying the optimal grouping strategy.

As established in Theorem~\ref{thm optimal group}, the optimal grouping consists of two types of antenna configurations: one group equipped with $L_1 = L_0 - r + 1$ antennas serving $t + L_1$ users, and the remaining $r - 1$ groups, each equipped with a single antenna and serving $t + 1$ users. Accordingly, the total number of users served per transmission slot is given by $g = (t+1)(r-1) + (t + L_0 - r + 1)$. To support this configuration, each user group must be allocated a disjoint set of antennas (i.e., column-wise separation) in every transmission slot to eliminate inter-group interference. Moreover, at most one group is permitted to use multiple antennas—up to $L_0 - r + 1$—while all other groups are restricted to a single active antenna per row.
This design allows one group to leverage spatial multiplexing through multi-antenna beamforming, while all groups—including the multi-antenna one—rely on RIS-based isolation to suppress inter-group interference. The resulting structure is consistent with the optimal configuration characterized in Theorem~\ref{thm optimal group}.

\begin{thm}
\label{th-PDA-RIS}	
Given a $(L_0, K, F, Z, r, S)$ RMAPDA, there exists a $(L, G, K, M, N)$ RIS-assisted multi-antenna coded caching scheme where $L = L_0$, $G = 2(r - 1)((t + 2)L_0 - r)$,  $M/N = Z/F$ and the   sum-DoF 
%$\min\{K, (F-Z)K/S\}$.
$(F-Z)K/S$. 
\end{thm}	

The underlying $F \times K$ array $\mathbf{P}$ consists of ``$*$'' symbols and integers, where each column corresponds to a user and each row represents a file subpacket. A ``$*$'' at position $(j, k)$ indicates that user $k$ caches the $j$-th subpacket of every file, while each integer $s \in [S]$ corresponds to a multicast transmission.
During the placement phase, users store all subpackets indicated by “$*$” in their respective columns. In the delivery phase, each $s$ identifies a group of users and subpackets, which are further partitioned into $r$ disjoint subarrays $\mathbf{P}^{(s)}_1, \dots, \mathbf{P}^{(s)}_r$. Each subarray corresponds to a user group served either by a single RIS-assisted antenna (if the row weights are at most one), or by multiple antennas with beamforming (up to $L_0 - r + 1$), following the optimal grouping strategy in Theorem~\ref{thm optimal group}.

This structure ensures interference-free parallel transmissions while satisfying the physical constraints of RIS-assisted systems. As in MAPDA, the cache-to-library ratio is given by $M/N = Z/F$, and the resulting scheme enables one-shot linear delivery with uncoded placement. 
% The complete delivery and placement procedures are summarized in Algorithm~\ref{alg:RMAPDAcoded}.

\begin{algorithm}
\caption{Caching Scheme Based on RMAPDA}
\begin{algorithmic}[1]
\Procedure{Placement}{$\mathbf{P}, \Wc$}
    \State Split each file $W_n \in \Wc$ into $F$ subpackets, i.e., $W_n = \{W_{n,f} \mid f \in [F]\}$.
    \For{$k \in [K]$}
        \State $Z_k \gets \{ W_{n,f} \mid \mathbf{P}(f,k) = *, \, n \in [N], \, f \in [F] \}$.
        \Comment{User $k$ caches subpacket $f$ of every file if $\mathbf{P}(f,k) = *$}
    \EndFor
\EndProcedure

\Procedure{Delivery}{$\mathbf{P}, \Wc, \mathbf{d}$}
    \For{$s \in [S]$}
        \State Identify subarrays $\mathbf{P}^{(s)}_1, \dots, \mathbf{P}^{(s)}_r$ satisfying RMAPDA condition C4;
        \For{$i \in [r]$}
            \State Determine the antenna group size $L_i$ for subarray $\mathbf{P}^{(s)}_i$;
            \If{$L_i = 1$}
                \State Transmit the multicast message to $t+1$ users using single-antenna transmission via RIS;
            \Else
                \State Transmit the multicast message to $t + L_i$ users using $L_i$-antenna beamforming;
            \EndIf
        \EndFor
    \EndFor
\EndProcedure
\end{algorithmic}
\label{alg:RMAPDAcoded}
\end{algorithm}

\subsection{Construction of RMAPDA}
% \subsection{Construction of RMAPDA}
In the following, we propose a new RMAPDA constructed based on the MN PDA in Lemma~\ref{lem:MNPDA} and the SMK MAPDA in Lemma~\ref{lem:SMK MAPDA}.

To construct the   RMAPDA, we employ the SMK MAPDA to serve users in the first group (i.e., the group containing $t+L_1$ users), while the MN PDA is used to serve users in each of the other groups (each containing $t+1$ users). However, the number of time slots in the SMK MAPDA and the MN PDA do not match. To resolve this, we replicate the SMK MAPDA and MN PDA multiple times to align their time slots. The resulting RMAPDA is formally described in the theorem below, with its detailed proof provided in Section~\ref{sec:bipartite scheme}. %and Appendix~\ref{sec:proof of bipartite RMAPDA}.

\begin{thm}[RMAPDA based on bipartite graph: $L_0 + tr \leq K$]
\label{thm:bipartite RMAPDA}
For any positive integers $K$, $L_0$, $r$ and $t = KM/N$ satisfying $ t<K$, $K \geq L_0+tr$ and $L_1=L_0-(r-1)>0$, 
there exists a $(L_0, K,F, Z, r, S = \alpha)$ RMAPDA,  where $Z= \alpha \left( \frac{1}{\binom{t+L_1-1}{t} \binom{K}{t+L_1}} + \frac{(r-1)}{\binom{K}{t+1}} \right) \binom{K-1}{t-1}$, $F= \alpha  \frac{(t+L_1) + (r-1)(t+1)}{K-t}$, $\lambda_1=\frac{\prod_{i=0}^{r-2} \binom{K - i(t+1)}{t+1}}{(r-1)!}$, and $\alpha=\operatorname{lcm}(\binom{t+L_1-1}{t} \binom{K}{t+L_1}, \lambda_1)$. Then for any integer $L\geq L_0$, we have a $(L,G = 2(r-1)((t+2)L_0-r),K,M,N)$ RIS-assisted multi-antenna coded caching scheme with the memory ratio $M/N=t/K$, the subpacketization $F$, and the sum-DoF  $L_0+tr$.
\end{thm}
 \begin{remark}
\label{remark:L_0 + tr > K}
When $L_0 + tr > K$, using the same construction method as in the case $L_0 + tr \leq K$, 
we can obtain a scheme with sum-DoF $K$. 
The detailed proof is included in Appendix \ref{appendix:RMAPDA $L_0 + tr > K$}.
\end{remark}

The RMAPDA constructed in Theorem~\ref{thm:bipartite RMAPDA} achieves the maximum sum-DoF as stated in Lemma~\ref{lem:MaxDoF}. 

\section{Proof of Theorem \ref{thm:bipartite RMAPDA}: New RMAPDA from bipartite graph}
\label{sec:bipartite scheme}
In this section, we present the detailed construction for Theorem~\ref{thm:bipartite RMAPDA} based on bipartite graph. Thus we first introduce the following notations on bipartite graph and its useful result.

A bipartite graph, denoted by $\mathcal{G}=(\mathcal{X},\mathcal{Y}; \mathcal{E})$, is a graph whose vertex set is partitioned into two disjoint subsets $\mathcal{X}$ and $\mathcal{Y}$, such that each edge in $\mathcal{E}$ connects a vertex in $\mathcal{X}$ to a vertex in $\mathcal{Y}$.  The degree of a vertex $v$ is the number of edges incident with it, denoted by $\text{d}(v)$. $\mathcal{G}$ is called regular if every vertex in $\mathcal{X}\cup \mathcal{Y}$ has the same degree.  When $|\mathcal{X}|=|\mathcal{Y}|$, a subset containing $|\mathcal{X}|$ edges $\mathcal{M} \subseteq \mathcal{E}$ is called a perfect matching if no two edges in $\mathcal{M}$ share a common vertex. Equivalently, a perfect matching establishes a one-to-one correspondence between 
the vertices of $\mathcal{X}$ and $\mathcal{Y}$.
The following lemma provides a sufficient condition for the existence of a perfect matching.
 \begin{lem}[Perfect matching \cite{zhang2020dynamic}]
Given a regular bipartite graph $\mathcal{G} = (\mathcal{X}, \mathcal{Y}; \mathcal{E})$ with $|\mathcal{X}| = |\mathcal{Y}|$, there exists a perfect matching between $\mathcal{X}$ and $\mathcal{Y}$.
\label{lem:matching}
\end{lem}

We now present an example to illustrate the construction strategy for the scheme in Theorem~\ref{thm:bipartite RMAPDA}.

\subsection{Example of Bipartite graph Scheme in Theorem~\ref{thm:bipartite RMAPDA}}
\label{sec:example}
\begin{figure*}[t]
    \centering    \includegraphics[width=0.8\linewidth,height=1\linewidth]{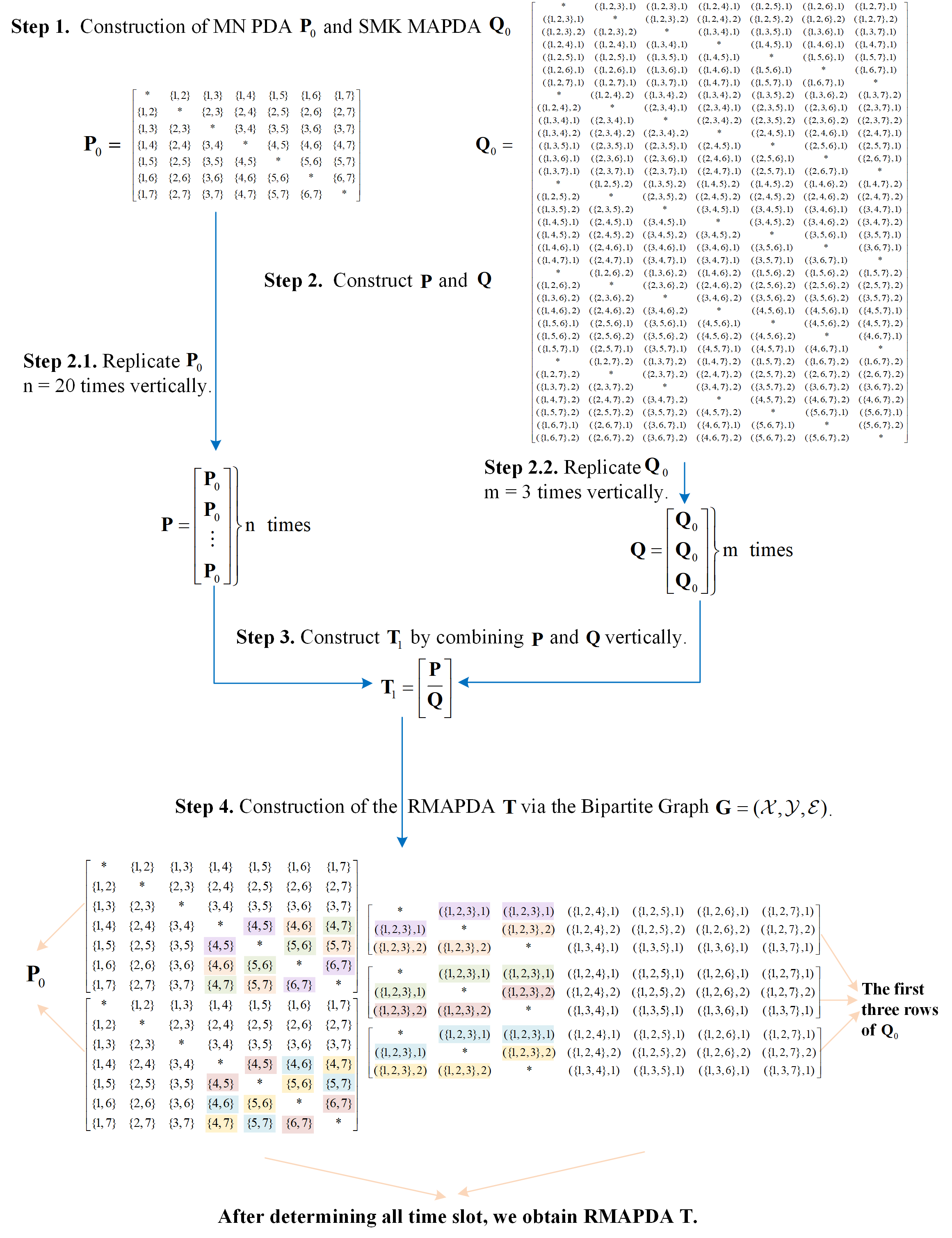}
    \caption{Construction of RMAPDA for Example.}
    \label{fig:example of general construction}
\end{figure*}

% The pseudocode of the proposed scheme is provided in Algorithm \ref{alg:RMAPDA}. 
 Consider a system with $K=7$, $L_0=4$, $M=1$, and $t = KM/N = 1$, where  %By the grouping method in Theorem~\ref{thm optimal group},   the antennas are divided into $r=3$ groups: one group with $L_1=2$ antennas and the other two groups each containing a single antenna. 
Fig.~\ref{fig:example of general construction} provides an overview of the construction process of RMAPDA for this example. 

  According to the optimal grouping strategy described in Theorem~\ref{thm optimal group}, all users and antennas are partitioned into $r =3$ groups in each time slot, where   the first group comprises $L_1=2$ antennas and each of the remaining $r-1=2$ groups    consists of a single antenna. 
In general,  we aim to design an RMAPDA, which is a combination of  two parts. The first part (corresponding to the  first group of antennas) is  constructed from the SMK MAPDA   for the MISO system with $L_1=2$ antennas; the second part (corresponding to the  second and third groups of antennas) is constructed from the MN PDA for the single-antenna system. 
In the following, based on the SMK MAPDA in Lemma~\ref{lem:SMK MAPDA} and the MN PDA in Lemma~\ref{lem:MNPDA}, we describe our construction through four steps. 
% \begin{figure}[H]
%     \centering
%     \includesvg[width=\linewidth]{6example_of_construction}
%     \caption{Construction of RMAPDA for Example.}
%     \label{fig:example of general construction}
% \end{figure}
\begin{itemize}
\item \textbf{Step 1. Constructing SMK MAPDA $\mathbf{Q}_0$ and MN PDA $\mathbf{P}_0$}: We first construct a $(7,7,1,21)$ MN PDA $\mathbf{P}_0$  in Table~\ref{tab:MN PDA example} and a $(2,7,35,5,70)$ SMK MAPDA $\mathbf{Q}_0$ in Table~\ref{tab:SMK MAPDA example} using Construction~\ref{const:MN PDA} and Construction~\ref{const:SMK MAPDA}, respectively. 
\item \textbf{Step 2. Constructing $\mathbf{P}$ and $\mathbf{Q}$}: Directly concatenating SMK MAPDA $\mathbf{Q}_0$ and MN PDA $\mathbf{P}_0$ typically results in a mismatch between the number of sets in the MN PDA and the number of sets in the SMK MAPDA. In order to ensure the alignment of  time slots, $\mathbf{P}_0$ and $\mathbf{Q}_0$ are respectively replicated to obtain the following two new arrays.
\begin{itemize}
\item \textbf{Step 2.1. Constructing $\mathbf{P}$}: For the MN PDA $\mathbf{P}_0$, in each time slot, we need to select two user groups $\mathcal{S}_1$ and $\mathcal{S}_2$, each containing two users, i.e., $\mathcal{S}_1, \mathcal{S}_2 \in \binom{[7]}{2}$, to be served simultaneously. Moreover, %to ensure that all possible combinations are accounted for, 
 we replicate $\mathbf{P}_0$ $n_1 = 10$ times so that every clique in $\mathbf{P}_0$ can be combined 
 with another two-user group; for example, 
 %partitioned into two column-disjoint cliques in all possible ways. For instance, the clique $\{4,5\}$. 
 all valid two-clique combinations that include $\{4,5\}$ are:
            \[
                \begin{aligned}
                &\{1,2\},\{4,5\}, \quad \{1,3\},\{4,5\}, \quad \{1,6\},\{4,5\}, \quad \{1,7\},\{4,5\}, \quad \{2,3\},\{4,5\}, \\
                &\{2,6\},\{4,5\}, \quad \{2,7\},\{4,5\}, \quad \{3,6\},\{4,5\}, \quad \{3,7\},\{4,5\}, \quad \{6,7\},\{4,5\},
            \end{aligned}
            \]
            giving a total of
            \[
            n_1 = \frac{\prod\limits_{i=1}^{r-2} \binom{K - i(t+1)}{t+1}}{(r-2)!} = \binom{5}{2} = 10
            \]
            such combinations. %Since each $\mathbf{P}_0$ can provide only one clique $\{4,5\}$, $\mathbf{P}_0$ needs to be replicated $n_1 = 10$ times. Each of $\mathbf{P}_0$ contains a total of $\binom{7}{2} = 21$ cliques. 
             After replicating $n_1 = 10$ times of $\mathbf{P}_0$, we obtain $\frac{21 \times 10}{2} = 105$ two-clique combinations, corresponding to 105 time slots. On the other hand, $\mathbf{Q}_0$ contains $\binom{7}{3}  \binom{2}{1} = 70$ vectors, which also correspond to 70 time slots. To match the total number of time slots between MN PDA and SMK MAPDA, $\mathbf{P}_0$ must be further replicated by an additional $n_2 = 2$ times on top of the initial 10 replications. Thus the $\mathbf{P}_0$ is replicated $n = n_1n_2 = 20$ times to get $\mathbf{P}$. 
            
 \item \textbf{Step 2.2. Constructing $\mathbf{Q}$}:  Based on the analysis in \textbf{Step 2.1}, to ensure that the number of time slots is equal, $\mathbf{Q}_0$ needs to be replicated $m = 3$ times vertically to get $\mathbf{Q}$.
\end{itemize}
 \item \textbf{Step 3. Construct $\mathbf{T}_1$ by combining $\mathbf{P}$ and $\mathbf{Q}$ vertically.}   It is not difficult to check that $\mathbf{T}_1$ has exactly $F=245$ rows and $Z=35$ stars in each column.
      
%        The final array $\mathbf{T}_1$, representing the overall cache placement scheme, is constructed by vertically concatenating the array $\mathbf{P}$ and the array $\mathbf{Q}$.
        \item \textbf{Step 4. Constructing the RMAPDA $\mathbf{T}$ via Bipartite Graph}: 
        After obtaining $\mathbf{T}_1$, we need to connect each time slot in $\mathbf{P}$ with a time slot in  $\mathbf{Q}$. 
        Based on $\mathbf{P}$ and $\mathbf{Q}$, we construct a bipartite graph $\mathcal{G}= (\mathcal{X}, \mathcal{Y}, \mathcal{E})$.  Each vertex in $\mathcal{X}$ corresponds to a vector in the $\mathbf{Q}$ (such as $(\{1,2,3\},1)$ in Fig.~\ref{fig:example of general construction}), and each vertex in $\mathcal{Y}$ corresponds to a 2-clique combination in the $\mathbf{P}$ (such as the combination of $\{4,5\}$ and $\{6,7\}$ colored in purple in Fig.~\ref{fig:example of general construction}). Each vertex $\mathcal{A} \in \mathcal{X}$ is connected to a vertex $\mathcal{S} = \{\mathcal{S}_1, \mathcal{S}_2\} \in \mathcal{Y}$ if and only if $\mathcal{A} \cap \mathcal{S}_i = \emptyset$ for all $i \in [2]$. Then we can check that the degree of each vertex in $\mathcal{X}$ is $2\times \frac{\binom{4}{2}\binom{2}{2}}{2!} = 6$ and in $\mathcal{Y}$ is $3\times \binom{3}{3}\binom{2}{1} = 6$. This implies that $\mathcal{G}$ is regular. 
        By Lemma~\ref{lem:matching} , there exists a perfect matching $\mathcal{M}$, ensuring a one-to-one correspondence between $\mathbf{Q}$ vectors and $\mathbf{P}$ clique combinations.
        
           For example,  the purple vector $(\{1,2,3\},1)$in SMK MAPDA and the 2-clique combination $\{\{4,5\},\{6,7\} \}$ in MN PDA, as shown in Fig.~\ref{fig:example of general construction}, form a pair of matched vertices in the bipartite graph $\mathcal{G}$, representing a single time slot. Fig.~\ref{fig:partial_bipartite_graph} illustrates some edges in a perfect matching $\mathcal{M}$ corresponding to the $6$ time slots marked with the same color in Fig.~\ref{fig:example of general construction}.
    %It corresponds to the connection patterns in the six colored time slots shown in Fig.~\ref{fig:example of general construction}. 
 \begin{figure}
                \centering
                \includegraphics[width=0.7\textwidth]{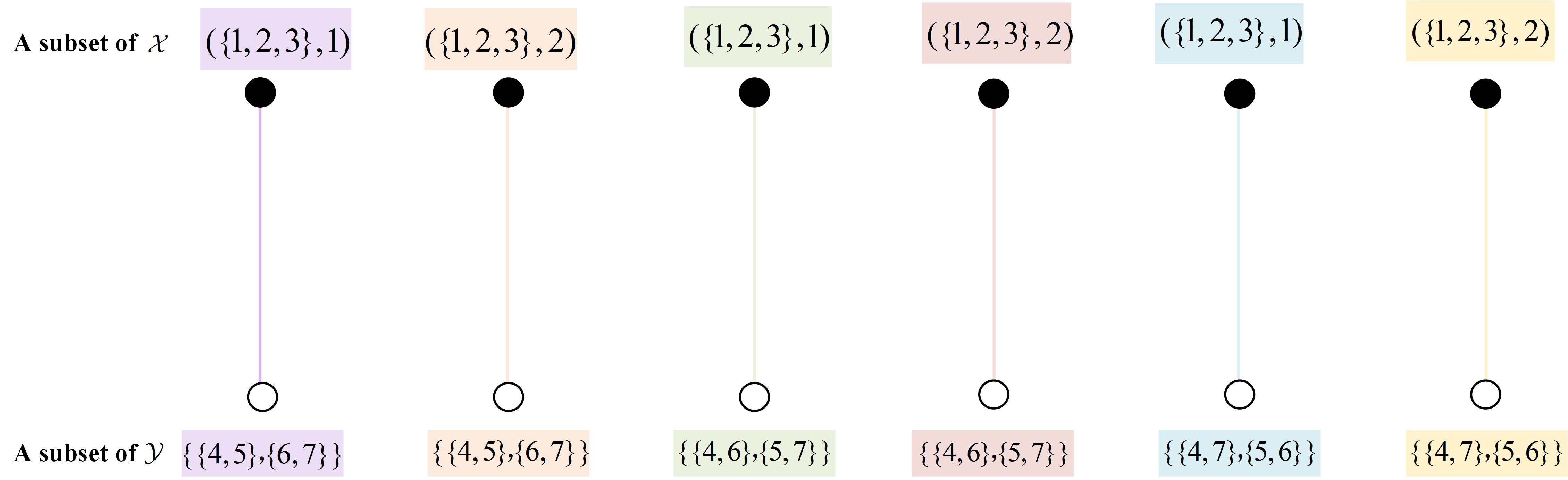}  
                \caption{
                    A bipartite subgraph illustrating the matched vertex pairings between subsets of \( \mathcal{X} \) and \( \mathcal{Y} \).       
                }
            \label{fig:partial_bipartite_graph}
            \end{figure}
There are exactly $S=210$ edges in the perfect matching. When each edge in $\mathcal{M}$ is regarded as a unique integer, the desired RMAPDA array $\mathbf{T}$, i.e. a $(4,7,245,35,3,210)$ RMAPDA, can be obtained. 
     
\end{itemize}

Based on the above $(4,7,245,35,3,210)$ RMAPDA, by Theorem \ref{th-PDA-RIS}, we can obtain a coded caching scheme for a $(4,36,7,1,7)$ RIS-assisted multi-antenna system where $M/N = Z/F = 1/7$ with sum-DoF $7$. 

\subsection{General Construction of Bipartite Graph-Based RMAPDA}
\label{sec:RMAPDA general}
    Consider a system with parameters $K$, $L_0$, and $M$, where $t = KM/N$. In each time slot, all users and antennas are divided into $r$ groups: the first group is equipped with $L_1$ antennas, while each of the remaining $r - 1$ groups contains a single antenna.     Our construction consists of four steps.
    \begin{itemize}
        \item \textbf{Step 1. Constructing SMK MAPDA $\mathbf{Q}_0$ and MN PDA $\mathbf{P}_0$.} 

        \item \textbf{Step 2. Construct $\mathbf{P}$ and $\mathbf{Q}$.}

        \begin{itemize}
            \item \textbf{Step 2.1. Construct  $\mathbf{P}$ by  replicating $\mathbf{P}_0$ $n$ times vertically, } where 
            \begin{align}
                n = n_1 n_2 = \frac{\prod\limits_{i=1}^{r-2} \binom{K - i(t+1)}{t+1}}{(r-2)!} \times \frac{\operatorname{lcm} \left( \binom{t+L_1-1}{t} \binom{K}{t+L_1}, \; \frac{\prod_{i=0}^{r-2} \binom{K - i(t+1)}{t+1}}{(r-1)!} \right)}{ \frac{\prod_{i=0}^{r-2} \binom{K - i(t+1)}{t+1}}{(r-1)!} }.
            \end{align}
            
            \item \textbf{Step 2.2. Construct  $\mathbf{Q}$ by  replicating $\mathbf{Q}_0$ $m$ times vertically,} where
            \begin{align}
                m =  \frac{\operatorname{lcm} \left( \binom{t+L_1-1}{t} \binom{K}{t+L_1}, \; \frac{\prod_{i=0}^{r-2} \binom{K - i(t+1)}{t+1}}{(r-1)!} \right)}{ \binom{t+L_1-1}{t} \binom{K}{t+L_1} }. 
            \end{align}
            \end{itemize}
                
        \item \textbf{Step 3. Construct $\mathbf{T}_1$ by combining $\mathbf{P}$ and $\mathbf{Q}$ vertically.} 

        \item \textbf{Step 4. Construction of the RMAPDA $\mathbf{T}$ via the Bipartite Graph.} In other words, after obtaining $\mathbf{T}_1$, we need to connect each time slot in $\mathbf{P}$ with a time slot in  $\mathbf{Q}$. This can be done as follows: 
        \begin{itemize}
            \item Based on the replicated arrays $\mathbf{P}$ and $\mathbf{Q}$, we construct a bipartite graph $\mathcal{G} = (\mathcal{X}, \mathcal{Y}, \mathcal{E})$.  Each vertex in $\mathcal{X}$ corresponds to a vector in the $\mathbf{Q}$, and each vertex in $\mathcal{Y}$ corresponds to a $(r-1)$-clique combination in the MN PDA . Each vertex $\mathcal{A} \in \mathcal{X}$ is connected to a vertex $\{ \mathcal{S}_1, \mathcal{S}_2, \ldots, \mathcal{S}_{r-1}\} \in \mathcal{Y}$ if and only if $\mathcal{A} \cap \mathcal{S}_i = \emptyset$ for all $i \in [r-1]$.

            \item  Based on the perfect matching in the bipartite graph $\mathcal{G}$, determine the user groups served in each time slot $s$, and replace the corresponding entries with the slot index $s$ to obtain the final RMAPDA array $\mathbf{T}$, i.e., a $(L_0,K,F,Z,r,S)$ RMAPDA. 
\end{itemize}
\end{itemize}

For the proposed RMAPDA construction, the following lemma will be proved in Appendix~\ref{sec:proof of bipartite RMAPDA}. 
\begin{lem}
\label{lem:lemma of PDA constraints}
The proposed RMAPDA %constructed in Theorem~\ref{thm:bipartite RMAPDA} 
construction based on a bipartite graph
satisfies all the conditions C1–C4 in Definition~\ref{defini:RMAPDA}.
\end{lem}

The  subpacketization level $F$ of this scheme is
\begin{align}
    \label{eq:total subpaketization}
    F &= m \binom{K-t-1}{L_1-1} \binom{K}{t} + n_1 n_2 \binom{K}{t}\\
    &= \frac{\operatorname{lcm}(\binom{t+L_1-1}{t} \binom{K}{t+L_1}, \frac{\prod_{i=0}^{r-2} \binom{K - i(t+1)}{t+1}}{(r-1)!})}{\binom{t+L_1-1}{t} \binom{K}{t+L_1}} \binom{K-t-1}{L_1-1} \binom{K}{t}\nonumber\\
    &+ \frac{\prod_{i=1}^{r-2} \binom{K - i(t+1)}{t+1}}{(r-2)!} \frac{\operatorname{lcm}(\binom{t+L_1-1}{t} \binom{K}{t+L_1}, \frac{\prod_{i=0}^{r-2} \binom{K - i(t+1)}{t+1}}{(r-1)!})}{\frac{\prod_{i=0}^{r-2} \binom{K - i(t+1)}{t+1}}{(r-1)!}} \binom{K}{t}\nonumber\\
    &= \alpha \cdot \frac{(t+L_1) + (r-1)(t+1)}{K-t}.
\end{align}
In addition, each user stores $Z$ packets of each file,  where
\begin{align}    
    Z &= \left( \frac{\operatorname{lcm}\left( \binom{t+L_1-1}{t} \binom{K}{t+L_1}, \frac{\prod_{i=0}^{r-2} \binom{K - i(t+1)}{t+1}}{(r-1)!} \right)}{\binom{t+L_1-1}{t} \binom{K}{t+L_1}} \right. \\
    &\quad + \left. \frac{\prod_{i=1}^{r-2} \binom{K - i(t+1)}{t+1}}{(r-2)!} \cdot \frac{\operatorname{lcm}\left( \binom{t+L_1-1}{t} \binom{K}{t+L_1}, \frac{\prod_{i=0}^{r-2} \binom{K - i(t+1)}{t+1}}{(r-1)!} \right)}{\frac{\prod_{i=0}^{r-2} \binom{K - i(t+1)}{t+1}}{(r-1)!}} \right) \binom{K-1}{t-1}\nonumber\\
    &= \alpha \left( \frac{1}{\binom{t+L_1-1}{t} \binom{K}{t+L_1}} + \frac{(r-1)}{\binom{K}{t+1}} \right) \binom{K-1}{t-1}.
\end{align}
Here, $\alpha=\operatorname{lcm}(\binom{t+L_1-1}{t} \binom{K}{t+L_1}, \lambda_1)$, $\lambda_1=\frac{\prod_{i=0}^{r-2} \binom{K - i(t+1)}{t+1}}{(r-1)!}$.

\section{Conclusion}
This paper introduced a new RIS-assisted multiple-antenna coded caching framework that leverages the interference nulling capability of RIS to enhance the system's multicast gain, quantified by the Degrees of Freedom (DoF). Unlike conventional multi-antenna coded caching models, our approach exploits a passive RIS with a limited number of elements to selectively suppress interference paths, thereby improving spectral efficiency. 
To achieve this, we proposed a new RIS-assisted interference nulling method that determines the phase-shift coefficients of the RIS with significantly faster convergence than existing algorithms. By partitioning transmissions into interference-free groups, where each group consists of a subset of antennas serving a subset of users, we reformulated the DoF maximization problem as a combinatorial optimization task. A low-complexity grouping algorithm was developed to obtain the optimal solution. %leading to the construction of a coded caching scheme that achieves the maximum DoF. 
Based on the grouping strategy, we introduced a new framework to construct RIS-assisted coded caching schemes, referred to as the \textit{RIS-assisted Multiple-Antenna Placement Delivery Array} (RMAPDA). Then a new bipartite graph-based RMAPDA construction achieving the optimal sum-DoF was proposed.
%and propose an efficient construction method for RMAPDA that significantly reduces subpacketization complexity. Simulation results validate the proposed framework, demonstrating substantial DoF improvements over existing schemes. 

Overall, this work highlighted the potential of RIS-assisted interference management in coded caching systems, providing a scalable and efficient solution for future wireless networks. Future research directions include exploring dynamic RIS configurations, optimizing practical deployment scenarios, and investigating the impact of imperfect channel state information on system performance.

\appendices  

\section{Proof of Theorem~\ref{thm optimal group}}
\label{sec:proof of thm optimal group}
 % \subsection{Given the number of active antennas}

Given the number of active antennas $L_0$, the total number of RIS elements $G$ must satisfy the following inequality:
\begin{align}
\frac{G}{2} &\geq \sum_{i=1}^r (L_i + t)(L_0 - L_i) \nonumber\\
&= \sum_{i=1}^r (L_0L_i + tL_0 - L_i^2 - tL_i) \nonumber\\
&= (L_0 - t) \sum_{i=1}^r L_i + trL_0 - \sum_{i=1}^r L_i^2 \nonumber\\
&= L_0(L_0 - t) + trL_0 - \sum_{i=1}^r L_i^2.
\label{eq:Ghalf}
\end{align}

To minimize the required RIS elements $G$, we aim to maximize the value of $\sum_{i=1}^r L_i^2$ under the constraint $\sum_{i=1}^r L_i = L_0$. Without loss of generality, we assume that the antenna allocation follows the ordering $L_1 \geq L_2 \geq \dots \geq L_r \geq 1$. 

Consider redistributing antennas from the second, third, $\dots$, and $r$-th groups to the first group. Suppose we move $a_2$, $a_3$, $\dots$, $a_r$ antennas, respectively, from groups $2$ to $r$ to the first group. The updated allocations are:
\begin{align*}
L'_1 &= L_1 + a_2 + a_3 + \dots + a_r, \\
L'_i &= L_i - a_i, \quad \forall 2 \leq i \leq r,
\end{align*}
where $a_i < L_i$ ensures that no group is entirely depleted. The updated sum of squared antenna allocations is:
\begin{align*}
\sum_{i=1}^r {L'_i}^2 &= (L_1 + a_2 + a_3 + \dots + a_r)^2 + (L_2 - a_2)^2 + (L_3 - a_3)^2 + \dots + (L_r - a_r)^2 \\
&= L_1^2 + L_2^2 + L_3^2 + \dots + L_r^2 + (a_2 + a_3 + \dots + a_r)^2 \\
&\quad + 2a_2(L_1 - L_2) + 2a_3(L_1 - L_3) + \dots + 2a_r(L_1 - L_r).
\end{align*}

Since $L_1 - L_2, L_1 - L_3, \dots, L_1 - L_r$ are all non-negative, we conclude that:
\begin{align*}
\sum_{i=1}^r {L'_i}^2 \geq \sum_{i=1}^r {L_i}^2.
\end{align*}

This implies that transferring more antennas to the group with the highest initial allocation increases $\sum_{i=1}^r {L_i}^2$. The extreme case occurs when the first group has exactly $L_0 - r + 1$ antennas, while each of the remaining $r - 1$ groups consists of a single antenna. This configuration maximizes $\sum_{i=1}^r {L_i}^2$, thereby minimizing the required RIS elements $G$ in \eqref{eq:Ghalf}.

\section{Proof of Proposition~\ref{prop:optimal_grouping_properties}}
\label{sec:Design Process of  alg:Optimal Grouping}
We first prove that given $g=L_0+ tr$, any feasible $L_0$ satisfies
    \[
    L_{\min} \le L_0 \le L_{\max},  %\quad L_0 \equiv g \pmod t,
    \]
    where $L_{\min}=\big\lceil g/(t+1)\big\rceil$ and $L_{\max}=\min\{L,\,g-t\}$.

Based on the analysis of scenarios where the number of active antennas is given, 
we can deduce that even if some antennas remain inactive, dividing the antennas into $r$ groups—
where $r-1$ groups each contain a single antenna and the remaining group consists of $L_0 - (r-1)$ antennas—
remains optimal under the grouping conditions.

To determine the scheme with the maximum number of antennas under a given sum-DoF $g$, 
we first establish the feasible range for $L_0$. 
The maximum achievable sum-DoF for $L_0$ antennas is $L_0(t+1)$, implying $L_0 \geq g/(t+1)$.
Since $L_0$ must be an integer, we refine this to $L_0 \ge \lceil g/(t+1) \rceil$.
From~\eqref{eq:dof}, we have
\begin{equation}
r = \frac{g - L_0}{t} \ge 1 \Rightarrow L_0 \le g - t.
\end{equation}
Thus, the range for $L_0$ is
\begin{equation}
    \lceil g / (t+1) \rceil \leq L_0 \leq \min\{L, g - t\},
\label{eq:the range of L_0}
\end{equation}
where $L$ is the total number of available antennas. 
In addition, $L_0$ must ensure $r=(g-L_0)/t$ is a positive integer.

Since $g = L_0 + t r$ and $r$ must 
be a positive integer, we have $g \bmod t = L_0 \bmod t$. Combined with the 
range $L_{\min} \le L_0 \le L_{\max}$, the two feasible endpoints inside this 
interval are
\[
L_0^{\max}=L_{\max}-(L_{\max}-g)\bmod t,\qquad
L_0^{\min}=L_{\min}+(g-L_{\min})\bmod t.
\]

We then prove that the optimal $L_0$ is either $L_0^{\max}$ or $L_0^{\min}$. 

Assume two schemes achieve the same sum-DoF $g$, with the number of active antennas in each scheme given by $L_0$ and $L_0'$, respectively, where $L_0 > L_0'$. 
Let the antennas be divided into $r$ and $r'$ groups, respectively. 
Given that
\begin{align*}
    g = L_0 + tr = L_0' + tr',
\end{align*}
it follows that $L_0 - L_0' = t(r' - r) > 0$, which implies $r < r'$. 
Let $\lambda = r' - r$.

From \eqref{eq:Ghalf}, the number of RIS elements required for $L_0$ antennas is
\begin{align}
\frac{G}{2} &= ((t+1)(L_0 -1) + (L_0 - (r-1) + t))(r-1) \nonumber \\
            &= (r-1)[(t+2)L_0 - r].
\label{eq:the initiation of G}
\end{align}
Similarly, for $L_0'$ antennas, the required number of RIS elements is
\begin{align}
\frac{G'}{2} &= ((t+1)(L_0' - 1) + (L_0' - (r'-1) + t))(r'-1) \nonumber \\
             &= (\lambda + r-1)[(t+2)(L_0 - \lambda t) - (\lambda + r)] \nonumber \\
             &= \lambda [(t+2)(L_0 - \lambda t) - (\lambda + r)] + (r-1)[(t+2)L_0 - r] \nonumber \\
             &\quad + (r-1)[-(t+2)\lambda t - \lambda].
\label{eq:changed G'}
\end{align}
Thus, the difference in the number of RIS elements between the two schemes is given by
\begin{align}
    f(\lambda) &= G' - G \nonumber \\
               &= 2\left[ \lambda ((-t^2 - 2t - 1)\lambda + (t+2)L_0 - 2r + (1-r)t^2 - 2tr + 2t + 1) \right].
\label{eq:diff}
\end{align}
Note that~\eqref{eq:diff} determines whether reducing the number of active antennas leads to an increase or decrease in the required number of RIS elements.
For a given sum-DoF $g$, let $L_0^{\max}$ denote the maximum number of antennas required to achieve this sum-DoF. 
Consider this as the reference scheme in \eqref{eq:diff}. 
By analyzing schemes with fewer antennas (i.e., different values of $\lambda$), we observe that $f(\lambda)$ is a downward-opening quadratic function passing through the origin with respect to $\lambda$. 
This implies:
\begin{itemize}
    \item If $f(\lambda) \geq 0$, reducing the number of antennas leads to an increase or no change in the number of RIS elements.
    \item If $f(\lambda) < 0$, using $L_0'$ antennas requires fewer RIS elements. A larger $\lambda$ results in a greater reduction in RIS elements.
\end{itemize}
Since $f(\lambda)$ is a concave quadratic function that passes through the origin at $\lambda = 0$ (i.e., $f(0)=0$ with a negative leading coefficient), its value first increases and then decreases as $\lambda$ grows. 
Consequently, when minimizing the total number of RIS elements 
\[
G(L_0',r') = G(L_0,r) + f(\lambda)
\]
over the discrete feasible lattice $\Lambda = \{0,1,\dots,\lambda_{\max}\}$, 
the minimizer of $G$ is equivalent to the minimizer of $f(\lambda)$ over $\Lambda$.  
Since $f(\lambda)$ is concave in $\lambda$, its minimum over the closed interval $[0,\lambda_{\max}]$—and thus over the discrete set $\Lambda$—must be attained at one of the two endpoints:
\[
\lambda^\star \in \{0,\ \lambda_{\max}\}.
\]
Here, $\lambda = 0$ corresponds to the scheme with the maximum number of active antennas $L_0^{\max}$, 
while $\lambda_{\max} \triangleq (L_0^{\max}-L_0^{\min})/t$ corresponds to the scheme with the minimum feasible number of active antennas $L_0^{\min} = L_0^{\max} - t\lambda_{\max}$. 

Hence, it suffices to evaluate $f(\lambda_{\max})$ (or equivalently, the two endpoint values of $L_0$) and select the configuration that yields the smaller RIS requirement:
\[
\text{if } f(\lambda_{\max}) < 0,\ \text{choose }\lambda = \lambda_{\max}; \quad
\text{otherwise, choose }\lambda = 0.
\]
Mapping $\lambda \in \{0, \lambda_{\max}\}$ back to $L_0$ coincides with the two endpoints in the feasible set. 
Therefore, the optimal $L_0$ is either $L_0^{\max}$ or $L_0^{\min}$. 

\section{Proof of Lemma~\ref{lem:lemma of PDA constraints}}
\label{sec:proof of bipartite RMAPDA}
Building on the illustrative example provided earlier, we now formally prove that the RMAPDA constructed via Theorem~\ref{thm:bipartite RMAPDA} satisfies Conditions C1–C4 defined for a valid RMAPDA.
    \begin{itemize}
    \item By the construction of the bipartite graph $\mathcal{G}$ proposed in \textbf{Step 4} in Section~\ref{sec:RMAPDA general}, we can verify that $\mathcal{X}$ and $\mathcal{Y}$ have  
 \begin{align}
	|\mathcal{X}| &= m \binom{t+L_1-1}{t} \binom{K}{t+L_1},\\
	|\mathcal{Y}| &= n_2 \frac{\binom{K}{t+1} \binom{K-(t+1)}{t+1} \cdots \binom{K-(r-2)(t+1)}{t+1}}{(r-1)!} = n_2\frac{\prod_{i=0}^{r-2} \binom{K - i(t+1)}{t+1}}{(r-1)!}
\end{align}vertices, respectively, and the degrees of each vertex in $\mathcal{X}$ and $\mathcal{Y}$ are 
 \begin{align}
 d_{\mathcal{X}} &= n_2\frac{\binom{K-(t+L_1)}{t+1} \binom{K-(t+L_1)-(t+1)}{t+1} \cdots \binom{K-(t+L_1)-(r-2)(t+1)}{t+1}}{(r-1)!},\label{eq:degree of X}\\
        d_{\mathcal{Y}}&= m\binom{K-(r-1)(t+1)}{t+L_1} \binom{t+L_1-1}{t},\label{eq:degree of Y}
    \end{align} respectively
where
\begin{align}
    m = \frac{\operatorname{lcm} \left( \binom{t+L_1-1}{t} \binom{K}{t+L_1}, \frac{\prod_{i=0}^{r-2} \binom{K - i(t+1)}{t+1}}{(r-1)!} \right)}{\binom{t+L_1-1}{t} \binom{K}{t+L_1}}.
    \label{eq:replication time for SMK MAPDA}
\end{align}
\begin{align}
    n_2 = \frac{\operatorname{lcm} \left( \binom{t+L_1-1}{t} \binom{K}{t+L_1}, \frac{\prod_{i=0}^{r-2} \binom{K - i(t+1)}{t+1}}{(r-1)!} \right)}{\frac{\prod_{i=0}^{r-2} \binom{K - i(t+1)}{t+1}}{(r-1)!}}.
\end{align}
\end{itemize}	
 After replication, we have 
\begin{align}
    |\mathcal{X}| = \operatorname{lcm} \left( \binom{t+L_1-1}{t} \binom{K}{t+L_1}, \frac{\prod_{i=0}^{r-2} \binom{K - i(t+1)}{t+1}}{(r-1)!} \right) = |\mathcal{Y}|; 
    \label{eq:numbers of vertices}
\end{align}
i.e., the time slot of SMK MAPDAS and the time slot of MN PDAs are matched.

For Condition C1, since the replication and vertical concatenation are performed based on the SMK MAPDA and MN PDA, and the number of ``$*$'' symbols in each column of the SMK MAPDA is identical to that in the MN PDA, the number of ``$*$'' symbols in each column of the resulting array remains unchanged after replication. Therefore, Condition C1 is satisfied.

For Condition C2, from the construction of the bipartite RMAPDA, it follows from~\eqref{eq:replication time for SMK MAPDA} and Construction~\ref{const:SMK MAPDA} that the number of time slots is given by
\begin{align*}
    S &= \binom{K}{t+L_1} \binom{t+L_1-1}{t} m \\
    &= \binom{K}{t+L_1} \binom{t+L_1-1}{t} \frac{\operatorname{lcm}(\binom{t+L_1-1}{t} \binom{K}{t+L_1}, \frac{\prod_{i=0}^{r-2} \binom{K - i(t+1)}{t+1}}{(r-1)!})}{\binom{t+L_1-1}{t} \binom{K}{t+L_1}}\\
    &= \operatorname{lcm}(\binom{t+L_1-1}{t} \binom{K}{t+L_1}, \frac{\prod_{i=0}^{r-2} \binom{K - i(t+1)}{t+1}}{(r-1)!}).
\end{align*}

For Condition C3, since it is inherently satisfied in the original SMK MAPDA and MN PDA before replication, the selection of users served in each time slot ensures that the columns corresponding to each user remain distinct. Consequently, Condition C3 is preserved after replication.

For Condition C4, since the users selected from the SMK MAPDA and the MN PDA are served simultaneously in the same time slot, it is sufficient to prove that the user groups in these two structures can be optimally matched. Specifically, for each time slot, the $t+L_1$ users in the SMK MAPDA and the $(r-1)(t+1)$ users in the MN PDA must be served simultaneously. If this condition holds, then Condition C4 is also satisfied.

To establish this, we employ the following Lemma~\ref{lem:matching} to demonstrate that the construction process ensures a perfect matching of the time slots in the SMK MAPDA and the MN PDA.

From~\eqref{eq:numbers of vertices}, we can see that 
\begin{align*}
    m \binom{t+L_1-1}{t} \binom{K}{t+L_1} = n_2 \frac{\binom{K}{t+1} \binom{K-(t+1)}{t+1} \cdots \binom{K-(r-2)(t+1)}{t+1}}{(r-1)!}.
\end{align*}
From previous discussions, we have a bipartite graph $\mathcal{G}=(\mathcal{X},\mathcal{Y},\mathcal{E})$. The degree of each vertex in $\mathcal{X}$ is given by:
\begin{align}
  d_{\mathcal{X}}
    &= n_2 \frac{\binom{K-(t+L_1)}{t+1} \binom{K-(t+L_1)-(t+1)}{t+1} \cdots \binom{K-(t+L_1)-(r-2)(t+1)}{t+1}}{(r-1)!}\nonumber\\
    &= n_2 \frac{[K-(t+L_1)]!}{[(t+1)!]^{r-1}(r-1)![K-(t+L_1)-(r-1)(t+1)]!}. 
    \label{eq:the degree of X}
\end{align}
Similarly, the degree of each vertex in  $\mathcal{Y}$ is given by:
\begin{align}
    d_{\mathcal{Y}} 
    &= m \binom{t+L_1-1}{t}  \binom{K-(r-1)(t+1)}{t+L_1} \nonumber\\
    &= m \binom{t+L_1-1}{t} \binom{K}{t+L_1} \frac{[K-(t+L_1)]! [K-(r-1)(t+1)]!}{K! [K-(r-1)(t+1)-(t+L_1)]!}\nonumber\\
    &= n_2 \frac{[K-(t+L_1)]!}{[(t+1)!]^{r-1}(r-1)![K-(t+L_1)-(r-1)(t+1)]!} \\
    &= d_{\mathcal{X}}.
    \label{eq:the degree of Y}
\end{align}

By combining~\eqref{eq:numbers of vertices}~and~\eqref{eq:the degree of Y} with Lemma~\ref{lem:matching}, it follows that the bipartite graph must have a perfect matching. This ensures that the multi-antenna and single-antenna groups can be perfectly matched, thereby satisfying Condition C4.

\section{Bipartite Graph RMAPDA for $L_0 + tr > K$}
\label{appendix:RMAPDA $L_0 + tr > K$}
In the following, we apply the bipartite-graph method in Theorem~\ref{thm:bipartite RMAPDA} to construct a scheme achieving a sum-DoF of $K$. When $L_0 + tr > K$—that is, when the system could in principle serve more users per slot than actually exist—the achievable sum-DoF cannot exceed $K$. In this regime, the number of users available in each slot is insufficient to form the optimal grouping structure. To preserve the framework in Section~\ref{sec:bipartite scheme}, which generates an RMAPDA by combining an SMK MAPDA with an MN PDA, we repartition the $K$ users and substitute the SMK component with a tailored $x$-MAPDA. Together with the MN PDA, and by following the same bipartite-graph construction in Section~\ref{sec:bipartite scheme}, the resulting RMAPDA continues to achieve the maximum sum-DoF of $K$ even under the condition $L_0 + tr > K$.

 The main idea is to choose integers $x,y$ such that
$K \;=\; x \;+\; y(t+1)$.
Because
$L_0+tr \;=\; \big(L_0-(r-1)+t\big) \;+\; (r-1)(t+1) \;>\; K$,
it follows that $x \le L_0-(r-1)+t$ and $y \le r-1$. We then construct two arrays: (i) a MADPA achieving sum-DoF $x$, and (ii) an MN PDA with coded-caching gain $t+1$. Using the bipartite-graph method, we schedule, in every time slot, the served users into $y$ disjoint groups of size $(t+1)$ and one additional group of size $x$, where $x = K - y(t+1)$. This yields an RMAPDA with per-slot grouping
consistent with $K = x + y(t+1)$ and, consequently, sum-DoF $K$. In what follows, we detail the construction according to the value of $L_0 + tr - K$.

\begin{itemize}
\item When $L_0 + tr-K < L_1$ where $L_1$ is defined in Theorem \ref{thm:bipartite RMAPDA}, we set $y = r - 1$ and $x = K - (r - 1)(t+1)$. 
We only use the $L_1'=x-t$  antennas to generate an SMK MAPDA with sum-DoF $(x-t)+t=x$ and use $r-1$ antennas for the MN PDA. Then by bipartite graph method, we can obtain our desired RMAPDA with sum-DoF $x+(r-1)(t+1)=K$.

\item When $L_0 +tr-K \geq L_1$,  let $y=\left\lfloor \frac{K}{t+1} \right\rfloor$ and $x = K - y(t+1)$. In this case, we have $x < t+1$. We will construct a special MAPDA (i.e., $x$-PDA) by transforming an MN PDA into a $x$-PDA in the following way. For convenience, we rewrite the construction of the MN PDA  $\mathbf{P}=(\mathbf{P}(\mathcal{T}, k))_{\mathcal{T}\in{\binom{[K]}{t}},k\in[K]}$ in Construction~\ref{const:MN PDA} as follows. For any $\mathcal{T}\in {\binom{[K]}{t}}$ and $k\in[K]$, the entry is defined as 
\begin{align}
            \mathbf{P}(\Tc, k) = 
        \begin{cases} 
        \ast & \text{if } k \in \Tc, \\
        (\mathcal{S}, \mathcal{S}^{-1}[k]) & \text{otherwise},
        \end{cases}
        \end{align}
        where $\mathcal{S} = \mathcal{T} \cup \{k\}$. Let $\mathcal{S}[i]$ denote the $i$-th smallest element of $\mathcal{S}$. For instance, if $\mathcal{S} = \{2,4\}$, then $\mathcal{S}[1] = 2$ and $\mathcal{S}[2] = 4$. Similarly, we define the inverse map $\mathcal{S}^{-1}$ such that $\mathcal{S}[i] = j$ if and only if $\mathcal{S}^{-1}[j] = i$.
        
 Then create $x$ vertically stacked copies $\{\mathbf{P}^{(0)}, \mathbf{P}^{(1)}, \dots, \mathbf{P}^{(x-1)}\}$ of $\mathbf{P}$.
We modify each vector $ (\mathcal{S}, \mathcal{S}^{-1}[k])$ of $\mathbf{P}$ to $ (\mathcal{S}, \lfloor\frac{\mathcal{S}^{-1}[k]-1+i(t+1)}{x}\rfloor)$. It is easy to check that the obtained array is a PDA with coded caching gain $x$. That is, we obtain a $x$-$(K,xF, xZ, S(t+1))$ PDA, which is also a MAPDA. Similar to the above case, we can also obtain an RMAPDA with sum-DoF $K$.
 
\end{itemize}
 
\bibliographystyle{IEEEtran}
\bibliography{reference}
\end{document}

%% file: macros.tex
\newcommand{\insertxfig}[4]{
\begin{figure}[htbp]
\begin{center}
\leavevmode \centerline{\resizebox{#4\textwidth}{!}{\input
#1.pstex_t}}
\caption{#2} \label{#3}
\end{center}
\end{fontsize}}

\long\def\comment#1{}

\newcommand{\Sc}{{\mathcal S}}
\newcommand{\Tc}{{\mathcal T}}

\newcommand{\Wc}{{\mathcal W}}

\newcommand{\Xc}{{\mathcal X}}

\renewcommand{\arg}{{\hbox{arg}}}

\newtheorem{thm}{Theorem}%[section]

\newtheorem{lem}{Lemma}

\providecommand{\definitionname}{Definition}